\documentclass[fleqn,usenatbib]{mnras}

\usepackage{newtxtext,newtxmath}

\usepackage[T1]{fontenc}
\DeclareRobustCommand{\VAN}[3]{#2}
\let\VANthebibliography\thebibliography
\def\thebibliography{\DeclareRobustCommand{\VAN}[3]{##3}\VANthebibliography}

\usepackage{graphicx}	
\usepackage{amsmath}	
\usepackage{scalerel}

\newcommand\HII{H\protect\scaleto{$II$}{1.2ex}}

\title[Scintillation of PSR~J0737$-$3039A]{Scintillating insights into PSR~J0737$-$3039A and \\ the interstellar plasma of the Gum Nebula from MeerKAT}

\author[Askew et al.]{
J.~Askew,$^{1,2}$\thanks{E-mail: jaskew@swin.edu.au}
D.~J.~Reardon,$^{1,2}$
R.~M.~Shannon,$^{1,2}$
M.~Bailes,$^{1,2}$
F.~Camilo,$^{3}$
A.~Corongiu,$^{4}$
M.~Kramer,$^{5,6}$
\newauthor
M.~E.~Lower,$^{7}$
A.~Parthasarathy,$^{5,8,9}$
A.~Possenti,$^{4}$ and
V.~Venkatraman~Krishnan,$^{5}$ 
\\
$^{1}$Centre for Astrophysics and Supercomputing, Swinburne University of Technology, P.O. Box 218, Hawthorn, Victoria 3122, Australia\\
$^{2}$Australian Research Council Centre of Excellence for Gravitational Wave Discovery (OzGrav)\\
$^{3}$South African Radio Astronomy Observatory, 2 Fir Street, Observatory 7925, South Africa\\
$^{4}$INAF-Osservatorio Astronomico di Cagliari, via della Scienza 5, 09047 Selargius (CA), Italy\\
$^{5}$Max-Planck-Institut f\"{u}r Radioastronomie, Auf dem H\"{u}gel 69, 53121, Bonn, Germany\\
$^{6}$Jodrell Bank Centre for Astrophysics, University of Manchester, Alan Turing Building, Manchester M13 9PL, UK\\
$^{7}$Australia Telescope National Facility, CSIRO, Space and Astronomy, PO Box 76, Epping, NSW 1710, Australia\\
$^{8}$ASTRON, Netherlands Institute for Radio Astronomy, Oude Hoogeveensedijk 4, 7991 PD Dwingeloo, The Netherlands\\
$^{9}$Anton Pannekoek Institute for Astronomy, University of Amsterdam, Science Park 904, 1098 XH Amsterdam, The Netherlands\\
}

\date{Accepted XXX. Received YYY; in original form ZZZ}

\pubyear{2025}

\begin{document}
\label{firstpage}
\pagerange{\pageref{firstpage}--\pageref{lastpage}}
\maketitle

\begin{abstract}
The double pulsar system PSR~J0737$-$3039A/B has enabled some of the most precise tests of strong-field gravity to date.
Here, we present a scintillation analysis of the system based on an 18-month observation campaign with the MeerKAT radio telescope.
The pulsar radiation shows flux density variations caused multi-path scattering, which results in an interference pattern that varies in frequency and time. 
We characterise this interference pattern to infer properties of scattering plasma and the orbital geometry of the system.
Our preferred model supports a scattering screen located at a distance of $D_s = 360^{+30}_{-40}$\,pc.
This moderately anisotropic screen of ionized gas (axial ratio $A_R = 2.4 \pm 0.2$) lies near the edge of the Gum Nebula, which is believed to be a supernova remnant (SNR) or an \HII\, region.
We estimate the expansion velocity of the nebula to be $V_{\textrm{s}} = 35 \pm 5$\,km\,s$^{-1}$, implying a SNR age of $t \approx 1$\,Myr.
We also constrain the orbital orientation and inclination sense of the double pulsar to be $\Omega = 40^{\circ} \pm 3^{\circ}$ and $i > 90^{\circ}$, respectively.
Assuming standard scattering geometry, our model yields a distance estimate consistent with the parallax-derived value of $D = 770 \pm 70$\,pc from very long baseline interferometry.
We conclude by discussing how future models of pulsar scintillation can enhance our understanding of the IISM and the properties of pulsars embedded within or lying behind such intervening structures.
\end{abstract}



\section{Introduction}\label{chapt:Introduction}

The pulsar timing technique has led to advancements in fundamental physics, in particular in understanding theories of gravity and general relativity \citep[GR; ][]{Kramer2005, Kramer2006, Kramer2021, Hu2022} and equation-of-state physics \citep[EoS; ][]{Lattimer2004, Antoniadis2016, Ozel2016, Lattimer2021, Hu2024}.
Breakthroughs have been made especially through the discovery of extreme relativistic binary systems and then studying their orbital orientations and component masses \citep{Hulse1975, Antoniadis2013, Cromartie2020, Kramer2021b}.
The best tests are made by studying extremely compact orbits with more massive companions \citep{Ozel2016, Kramer2021b}.
The sensitivity and precision of these tests also improve as the number of observations and timing baselines increase.
An excellent example of this is the only known double pulsar system, PSR~J0737$-$3039A/B.
The two pulsars (A and B) have rotational periods of 23\,ms and 2.8\,s, respectively \citep{Lyne2004}.
Not only have both neutron stars been detectable, they are in a compact 2.45\,hr orbit.
This allows for tests of GR with exquisite precision \citep{Kramer2006}.

The mass of components of a binary system can be measured using two post-Keplerian (PK) parameters: the range and shape of the Shapiro delay \citep{Shapiro1964}.
This measurement is sensitive to the orbital inclination of the system $i$, with more `edge-on' systems having stronger constraints on companion mass \citep{Camilo1994, Shamohammadi2023}.
For the double pulsar, the sense of this orientation has not been resolved, with early studies finding $i=88\rlap{.}^\circ1 \pm 0\rlap{.}^\circ5$ \citep{Rickett2014} but more recent work measuring $i=90\rlap{.}^\circ65\pm0\rlap{.}^\circ05$ \citep{Kramer2021, Hu2022, Lower2024}.
Determining the geometry and orientation of the double pulsar orbit is important for understanding stellar binary population synthesis models \citep{Stairs2006, Tauris2017, Riley2022}, establishing astrophysically motivated priors for kilonova \citep{Zhu2020}, and calculating when the radio beam of pulsar B will be detectable again \citep{Lyutikov2005, Breton2009, Perera2012, Noutsos2020}. 

The double pulsar system is also important because it is possible to measure seven PK parameters in the system.\footnote{This has enabled incredibly precise mass measurements of the system \citep[$\approx 0.5\,M_{\earth}$; ][]{Hu2022}.}
The most precise test of gravity comes from the emission of gravitational waves due to orbital period decay \citep{Hulse1975} and is consistent with GR at the 1.3 $\times$ 10$^{-4}$ (95\% confidence) level \citep{Kramer2021}.
A systematic error in the orbital period decay is introduced due to the correction for the Shklovskii effect \citep{Shklovskii1970}.
The error is caused by the uncertainty in the distance to the double pulsar. 
Although this error is negligible in current tests of gravity \citep{Kramer2021}, it will become more significant as timing precision improves \citep{Hu2020}.
Currently, measurements in the distance to the system from pulsar timing ($D=465^{+134}_{-85}$\,pc) and very long baseline interferometry (VLBI, $D=770\pm70$\,pc) are in slight tension with each other \citep{Kramer2021}.
Both of these methods are purely geometric.
The VLBI distance is determined from the semi-annual variations in the apparent position of the image of the pulsar \citep{Gwinn1986, Ding2023}.
The timing distance is inferred from semi-annual variations in pulse arrival times caused by the curvature of the radio wavefront as it passes the orbit of the Earth \citep{Kaspi1994}.
It is not uncommon to find differences between the distances determined by pulsar timing and VLBI, potentially caused by several systematic biases \citep{Lutz1973, Verbiest2012, Ding2023}.

This tension could be resolved by introducing a third independent distance measurement by studying the intensity variations of the scintillation pattern.
These variations are caused by scattering from the ionized interstellar medium (IISM) \citep{Rickett1968}.

The scattering along the line of sight is often dominated by a particular region of ionized plasma, which can then be approximated as a single dominant screen \citep{Sutton1971}.
The propagation of radio waves travelling along different paths results in constructive and destructive interference and variations in intensity in time and frequency.  
Bright patches in the pattern are known as `scintles'.
The properties of the scintillation pattern depend on the velocity, structure, and relative distance to this scattering screen \citep{Rickett1990, Cordes1998}.
When modelled and accounted for, these properties can be used to infer the distance to a pulsar \citep{Cordes1998, Mall2022, Reardon2024b}.
This model is dependent on the effective velocity along the line of sight, which is a weighted sum of the Earth, scattering screen, and pulsar velocities.
The transverse pulsar velocity component is sensitive to the orientation of the orbit and therefore resolves the degeneracy in the sense of the inclination angle from pulsar timing \citep{Ord2002, Rickett2014}.

We can use also use models of scattering screens to study the velocity, structure, density, and other properties of ionized gas along the line of sight \citep{Mall2022, Askew2023, Ocker2024}.
These techniques have been applied recently for several structures including those in our local IISM \cite[e.g., within 1\,kpc,][]{Mall2022, Ocker2024}, with a focus on the Local Bubble \citep[an under-dense region in which the Solar system resides,][]{Stinebring2022, Liu2023, ONeill2024, Stock2024}, the Loop I bubble \citep{Bhat1998}, star-forming nebulae \citep{Gupta1994, Mall2022}, hot (type O-A) stars  \citep{Walker2017}, pulsar wind nebulae \citep{Main2021}, and ionized supernova remnant (SNR) shells \citep{Yao2021}.

\begin{figure}
	\includegraphics[width=1\columnwidth]{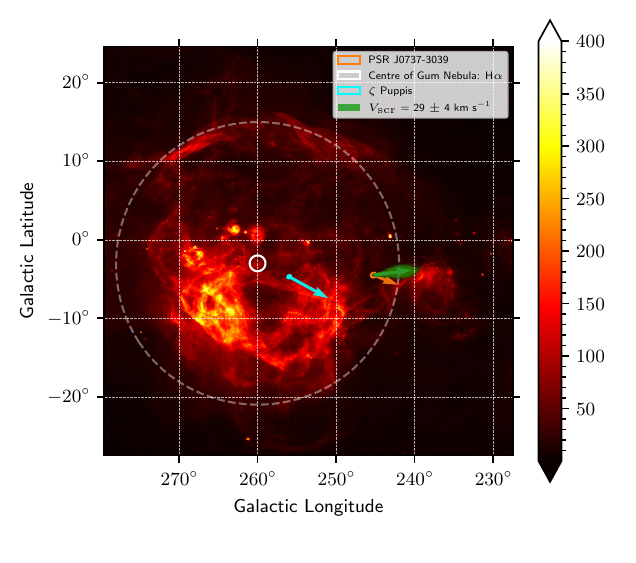} 
    \caption{ H$\alpha$ emission towards the Gum Nebula and PSR J0737$-$3039. The colour bar shows the intensity for a given Galactic latitude (y-axis) and longitude (x-axis) in units of ergs\,cm$^{-2}$\,sec$^{-1}$. The white 36$^{\circ}$-wide dashed circle is centred on the putative centre of the nebula. The cyan ellipse and arrow highlight the star $\zeta$ Puppis and its proper motion. The orange ellipse and vector show the location and proper motion of the double pulsar. The green vectors show the screen velocity from our model. The semi-transparent vectors are drawn from the posterior probability distribution to show the uncertainty in the velocity. The emission-line map was produced from the Southern H$\alpha$ Sky Survey Atlas \citep{Gaustad2001}. }
    \label{fig:Gum}
\end{figure}

The Gum Nebula is a noteworthy structure in the Milky Way that has been studied across the electromagnetic spectrum \citep{Gum1952, Reynolds1976A, Woermann2001, Sushch2011, ONeill2024, Wang2025}.
It is a complex mixture of overlapping regions of ionized and neutral material that appears close on the sky to the double pulsar (see Figure \ref{fig:Gum}).
The Gum Nebula has a large radius of 18$^{\circ}$, centred at approximately RA 08:22:53.6 and DEC $-$42:25:20 (J2000).
The distance to the Gum Nebula is still debated, with a range of estimates from $300$ to $500$\,pc \citep{Woermann2000, Woermann2001, Sushch2011, Pagani2012, Purcell2015, ONeill2024}.
The nebula is thought to have originated as a SNR or \HII\, region \citep{Gum1952, Reynolds1976A, Woermann2000, Sushch2011, Purcell2015, ONeill2024}.
If a scattering screen is situated at the edge of the Gum Nebula we could use it to measure the expansion and infer its age, density, and metallicity \citep{Cioffi1988, Sushch2011}.

This work is the result of an 18-month observing campaign of the double pulsar using the Meer Karoo Array Telescope (MeerKAT).
This campaign allowed us to study the properties of the intervening IISM, determine the inclination angle of the binary orbit, and investigate the distance to the system.
We describe the data collected with the MeerKAT radio telescope in Section \ref{chapt:Data}.
We discuss how we modelled the scintillation, and some of the challenges in modelling in Section \ref{chapt:Methods}.
The results of this modelling is presented in Section \ref{chapt:Results}.
We discuss the implications of our measurements of the orientation of and distance to the system in Section \ref{chapt:Discussion}.
In this section, we also present what we can understand about the ionized gas and physics of the Gum Nebula from our observations. 
Finally, in Section \ref{chapt:Conclusions} we present our main conclusions and discusses ongoing observations of the double pulsar.


\section{Data}\label{chapt:Data}

\subsection{MeerKAT Observations}\label{chapt:Observations}

This work uses data from MeerKAT, a radio interferometer comprising 64, 13.9-m antennas located in the Northern Cape province of South Africa \citep{MeerKAT2016}.
While observations of the double pulsar system have been ongoing since April 2019, the observations used in this analysis extend from June 2022 to November 2023. The observations were taken under the Relativistic Binary programme \citep[RelBin,][]{Kramer2021b}, part of the MeerTime Large Survey Project \citep{Bailes2020}.
MeerTime has many aims.
These aims include testing theories of gravity, probing neutron star masses in compact binary systems, and using pulsar timing observations to detect nanohertz frequency gravitational waves \citep{Spiewak2022, Gitika2023, Miles2023, Miles2025}.

MeerTime data are recorded using four independent Pulsar Timing User Supplied Equipment (PTUSE) signal processors, which ingest tied-array (voltage) beams produced by the MeerKAT correlator/beamformer (CBF) \citep{Bailes2020}.
The CBF produces voltage streams with configurable channelisation.
For most pulsar timing projects, a 1024-channel mode is used as it provides lower data rates than other modes, and is designed with a polyphase filter that reduces aliasing artefacts between channels.
The PTUSE machines are used in pulsar timing projects to record filterbank (search mode) and fold mode data for searching and timing, respectively \cite[see][for further details]{Bailes2020}.
Our observations were recorded in three observing bands: the UHF-band, L-band, and S-band, which are described in Table \ref{tab:bands}.
To resolve the scintillation in the UHF and L-band observations it is necessary to produce spectra with more than 1024 channels.
To accomplish this, in addition to recording data suitable for precision timing, we recorded data simultaneously using a second PTUSE machine.
The second PTUSE machine was configured to subdivide each of the 1024 coarse channels by a factor of 16, resulting in 16384 channels.
We refer to this data product as the 16\,k mode. 

\begin{table}
	\centering
	\caption{ Properties of the observing bands. Our analysis used 12 observations with the UHF receiver and two observations each using the L-band and S-band receivers.}
	\label{tab:bands}
	\begin{tabular}{l | c c r}
		\hline
		\hline
	    Observing Band & UHF & L-band & S-band \\
    	\hline
        Central Frequency (MHz) & 815.73 & 1283.58 & 2187.07 \\
        Bandwidth (MHz) & 544 & 856 & 875 \\
        Spectral Resolution (kHz) & 33 & 52 & 853 \\
        No. of Channels & 16384 & 16384 & 1024 \\    
        \hline
	\end{tabular}
\end{table}

For the double pulsar system, the A pulsar (PSR J0737$-$3039A) is observed by the RelBin theme every month for $\sim$3\,hr, which is modestly longer than an entire orbit.
This was chosen to deliver two eclipses during the observing campaign, in which the magnetosphere of pulsar B interferes with the radio emission from pulsar A \citep{Lower2024}.
The observations were divided into three consecutive scans of duration $\sim$30\,min, $\sim$2\,hr, and $\sim$30\,min respectively.

\subsection{Dynamic Spectra}\label{chapt:dynspec}

The fundamental data products used in this study are dynamic spectra.
We first produced these dynamic spectra from MeerPIPE, the pulsar timing data reduction pipeline, which utilises \texttt{PSRCHIVE} pulsar-data processing utility \citep{Hotan2004}.
Polarization and flux calibration are also performed on the data during this stage. 
This calibration is further described in \citep[][]{Serylak2021}.
To determine the flux density for each channel and each subintegration, we used the \texttt{psrflux} tool, which uses a standard pulse profile derived from pulsar timing as a match filter to measure the flux density with near optimal signal to noise ratio. 

\begin{figure*}
	\includegraphics[width=0.8\textwidth]{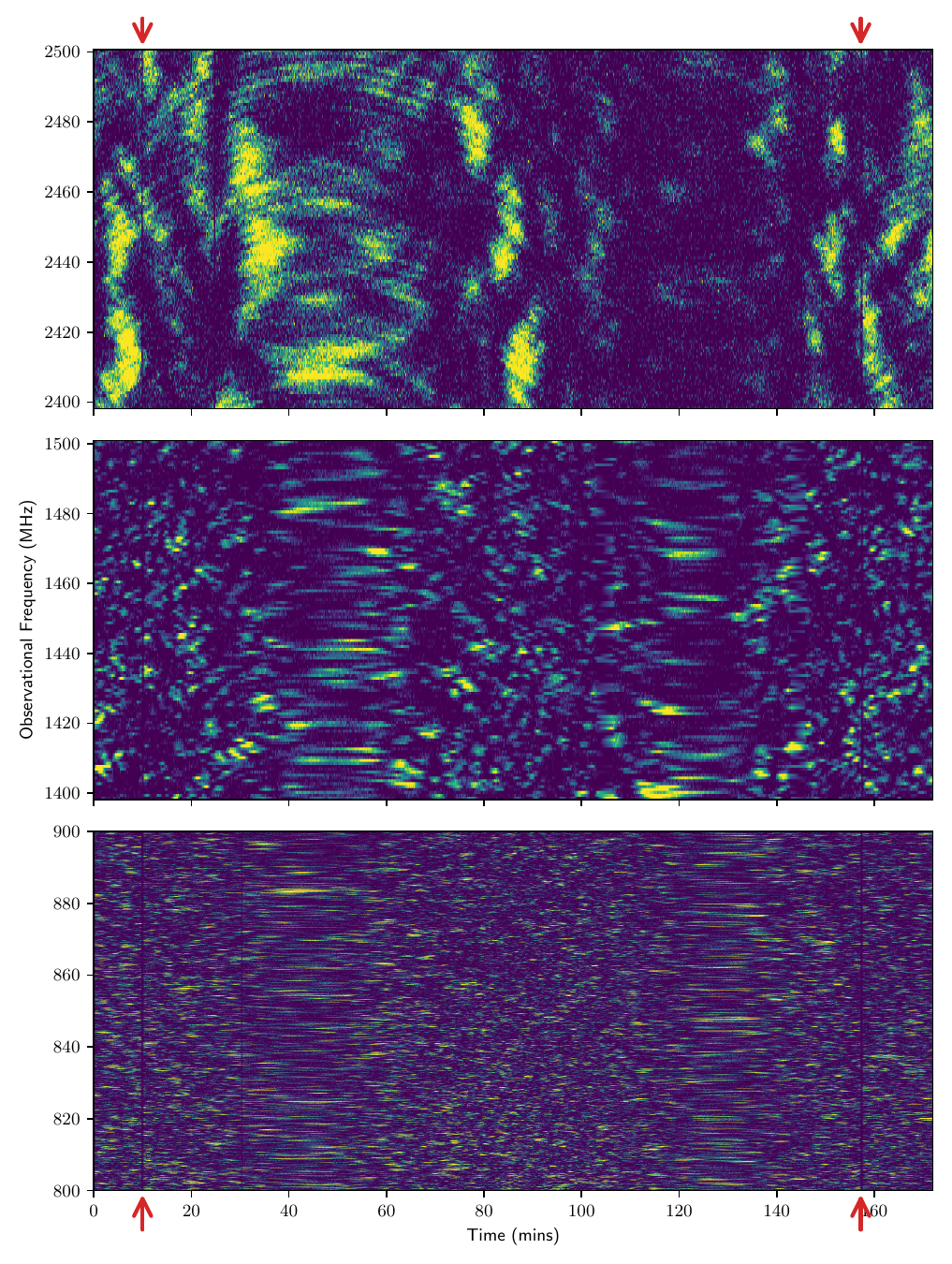} 
    \caption{ Dynamic spectra of PSR J0737$-$3039A  in S-band (top), L-band (middle), and UHF-band (bottom). Two eclipses, where the beam of pulsar A passes through the magnetosphere of pulsar B, can be seen, marked by the red arrows. The colour scales are all independent across each panel and use linear scaling. 
    The data presented in each panel show a 100\,MHz subband. Throughout the work we used the full bandwidth of each observation.}
    \label{fig:dynspec}
\end{figure*}

The main features of the scintillation pattern in the dynamic spectra are `scintles', which are islands of increased flux density that vary with frequency and time.
The decorrelation bandwidth of the scintles rapidly increases with frequency, as seen in Figure \ref{fig:dynspec}. 
The duration of the scintles changes with time as the transverse velocity of the pulsar changes across the orbit \citep{Rickett2014, Reardon2019}.

\section{Methods}\label{chapt:Methods}

Our analysis uses dynamic spectra taken across 16 epochs spanning 18 months in all three frequency bands.
In earlier observations (exclusively obtained at UHF and L band), the 16\,k mode was unavailable and therefore the scintillation bandwidth was not resolved in the observations.
The dynamic spectra were first divided into sub-dynamic spectra (which we term `slices') of smaller bandwidth and observing time to capture the rapidly changing scintillation bandwidth, $\Delta\nu_d$, (with frequency) and timescale, $\Delta\tau_d$, (with orbital phase).

Rapidly varying scintillation timescale is visually apparent  in the dynamic spectra (Figure \ref{fig:dynspec}). 
This motivates the division of the dynamic spectra into windows of time.
We chose to divide the windows into 10\,min durations as this is longer than the largest measured timescales, as listed in  Table \ref{tab:scintparams}, while providing uniform sampling across the orbital phase.
Windows of different lengths were tested, but provided over-estimation or under-estimation of the scintillation timescale.

\begin{table}
	\centering
	\caption{ The minimum, mean, and maximum scintillation bandwidths and timescales for each observing band.}
	\label{tab:scintparams}
	\begin{tabular}{l | c c r}
		\hline
		\hline
	    Observing Band & UHF & L-band & S-band \\
    	\hline
        Minimum $\Delta\nu_d$ (MHz) & 0.036 & 0.143 & 0.970 \\
        Mean $\Delta\nu_d$ (MHz) & 0.070 & 0.496 & 3.139 \\
        Maximum $\Delta\nu_d$ (MHz) & 0.261 & 1.073 & 7.876 \\
        \hline
        Minimum $\Delta\tau_d$ (s) & 36 & 46 & 51 \\
        Mean $\Delta\tau_d$ (s) & 63 & 80 & 130 \\ 
        Maximum $\Delta\tau_d$ (s) & 408 & 410 & 582 \\    
        \hline
	\end{tabular}
\end{table}

\begin{figure*}
	\includegraphics[width=1.5\columnwidth]{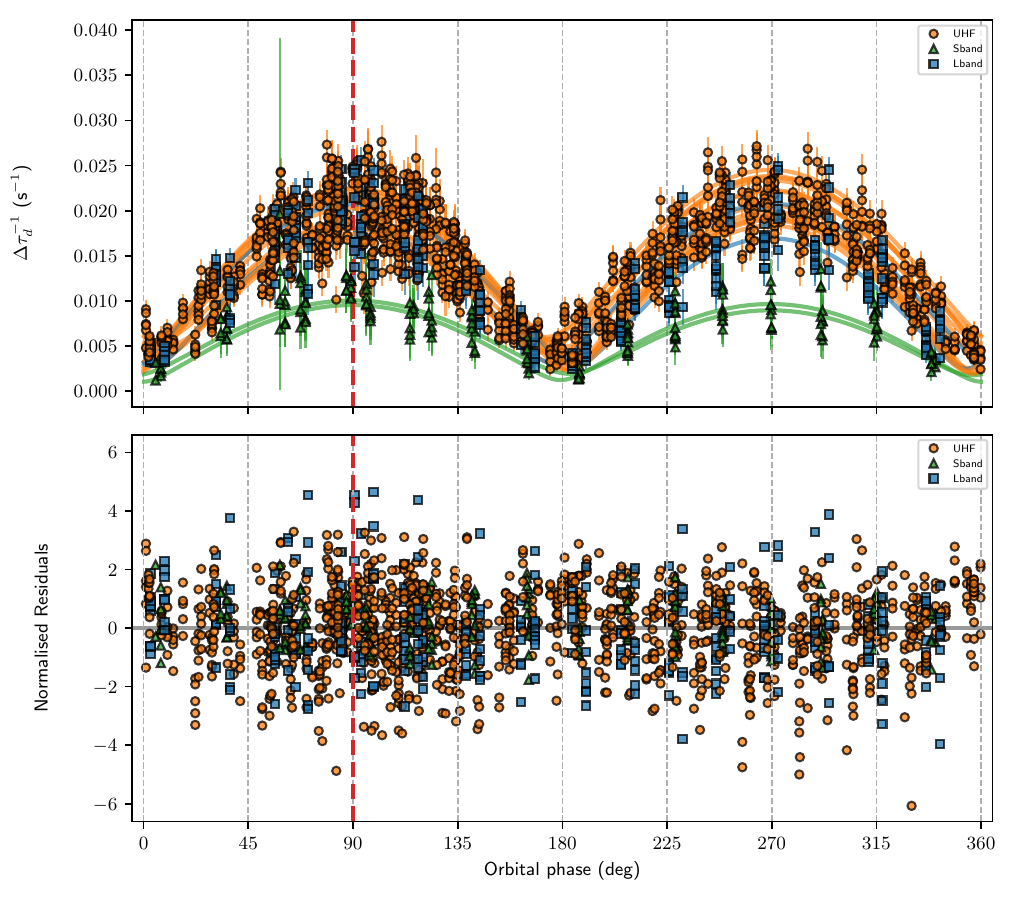}
    \caption{ Measurements of, and model for, $\Delta\tau_d^{-1}$. Top panel: Reciprocal of scintillation timescales for UHF (circles), L-band (squares), and S-band (triangles). The solid lines indicate the best-fitting model for $\textbf{V}_{\textrm{eff}}$ each epoch (16 in total). The maximum a-posteriori parameters for this model are presented in Table \ref{tab:table_model}. Lower panel: Normalised residuals for the best-fitting model. These are the differences between the observations and the model, divided by measured uncertainties. We discuss the whiteness of the residuals in Appendix \ref{chapt:AppendixB}. The dashed vertical line indicates the epoch of superior conjunction. Both panels are plotted against the orbital phase of the double pulsar, determined from the true anomaly expressed in degrees \citep[][Equation A4]{Reardon2019}.}
    \label{fig:taumodel}
\end{figure*}

\begin{figure}
	\includegraphics[width=1\columnwidth]{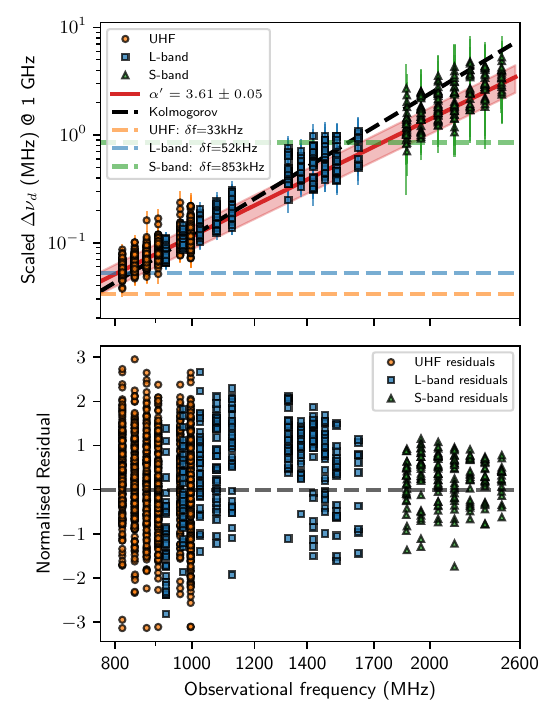}
    \caption{ Frequency dependence of scintillation for the double pulsar. The scintillation bandwidth measurements are shown at each epoch in the UHF (circles), L-band (squares), and S-band (triangles) observations. At each epoch, we infer a value for $\Delta\nu_{\textrm{d, 1GHz}}$. We rescale each epoch with the same median value to account for epoch-to-epoch variations in the scintillation bandwidth. The solid (red) line indicates the globally inferred value for $\alpha^{\prime}$ and its uncertainty accounting for the variation of $\Delta\nu_{\textrm{d, 1GHz}}$ between epochs. We also show a global model assuming Kolmogorov turbulence in black. The dashed lines show the channel bandwidth in each observing band. The second panel shows the normalised residuals of this scaled bandwidth (data - model / uncertainty).
    }
    \label{fig:alpha_prime}
\end{figure}

In addition, the division of the dynamic spectrum in frequency was necessary due to the large available fractional bandwidth.
The difference between the scintillation bandwidth at the top of each band compared to the bottom is $\sim$an order of magnitude (see Figure \ref{fig:dynspec}).
At each observing band there is also a large amount of scatter in the measurements of $\Delta\nu_d$ used in the analysis (see Table \ref{tab:scintparams} and Figure \ref{fig:alpha_prime}).
To account for the evolution of the frequency bandwidth at different bands, the UHF, L-band, and S-band observations were divided into 30\,MHz, 50\,MHz, and 100\,MHz sub-bands, respectively.
Some sub-bands were not included in the analysis due to corruption from radio frequency interference (RFI). 
Notably, the UHF band is nearly devoid of RFI at MeerKAT \citep{Bailes2020}. 
We only needed to consider removing one sub-band from the data analysis; and even in one case that sub-band was sufficiently clean that it could be used. 
Measurements of the scintillation properties were thus made in slices comprising sub-bands and windows. 
The scintillation bandwidth measurements are shown in Figure \ref{fig:alpha_prime}.

At the lowest frequencies observed in UHF (typically below 800\,MHz), the scintillation bandwidth is smaller than our 0.033\,MHz spectral resolution.
These frequencies were not used in the analysis.
In the S-band observations, the scintillation bandwidth was larger than the channel bandwidth in the standard 1024-channel mode, so we did not need to use the 16k-mode dynamic spectra.

\subsection{Scintillation Bandwidths and Timescales}\label{chapt:Scintparams}

For each slice, we use a two-dimensional (2D) auto-correlation function (ACF) to characterise the scintillation properties. 
In addition to measuring the scintillation time scale and bandwidth, which measure the bandwidth of the autocorrelation function in time and frequency, we also measure the tilt of the ACF. This is the rotation of the major axis of the ACG ellipse from the time or frequency axes, which can be caused by a large refractive phase gradient in the ISM \cite[][]{Rickett2014}. 
The rotation is quantified by a gradient $\phi$.
The 2D ACF is modelled using an analytical function parameterized by $\Delta\nu_d$, $\Delta\tau_d$, and $\phi$ \citep{Reardon2023}.
Following convention, $\Delta\nu_d$ is defined as the frequency lag at which the correlation falls to one-half its value and $\Delta\tau_d$ is the time lag at which the correlation falls to 1/$e$ of the peak value.
We measure these properties using the \texttt{SCINTOOLS}\footnote{\url{https://github.com/danielreardon/scintools}} \citep{Scintools} software package.
A more detailed description of this fitting method can be found in \citet{Rickett2014} and \citet{Reardon2023}.

During the analysis of the auto-correlation functions, we identified structures that needed to be accounted for prior to making robust estimation of scintillation timescales and bandwidths.
Firstly, we identified that the overestimation was caused by an artefact related to the polyphase filters only apparent in the 16-k data.
The artefact is an increase in covariance in the 2D$-$ACFs  at frequency lags 0.520\,MHz and 0.832\,MHz for UHF and L-band (in the 16\,k mode) respectively.
This is caused by the polyphase filters in the CBF used to divide the data in 1024 channels.
The filters have significant roll-off in sensitivity and signal close to channel edges, introduced to reduce spectral leakage between channels \citep[][]{Bailes2020}.
This artefact can cause an overestimation of $\Delta\nu_d$ by $\geq 16$\% (at UHF).

We also noticed diffuse structures in the ACF: excess signal at high frequency and time lags.
We attribute this to a second distinct scintillation pattern.
This diffuse structure could be caused by a second scattering screen .
In many cases, the phase gradience measured for the second component of the ACF is different (and occasionally orthogonal to) the primary component, which suggests the power is caused by a second screen.
Unfortunately, the excess power is too faint to robustly measure scintillation bandwidths and timescales.

To avoid incorrectly biasing our measurements to the presence of the filter artefact or a second scattering screen, we restrict the size of the ACF when performing a fit.
In the UHF band, the scintillation bandwidth $\Delta\nu_d$ is smaller ($< 0.2$\,MHz) than the artefact (0.52\,MHz).
In L-band we use the 1024 channel data to measure $\Delta\nu_d$ where $\Delta\nu_d > 0.832$\,MHz and the 16-k data when the scintillation bandwidth is smaller. 
This results in robust measurements of $\Delta\nu_d$ and $\Delta\tau_d$.

There is also a trade-off between the duration of the window of the dynamic spectra in time and frequency as larger bins lead to more scintles and therefore precise measurement.
This trade-off is caused by the `finite scintle effect': the ability to precisely measure scintillation bandwidths and timescales depends on the number of scintles in a dynamic spectrum \citep[][]{Cordes1986}.
In additional to including standard measurement error, the uncertainties in the scintillation bandwidth and time scale both includes terms that are inversely proportional to the square root of the number of scintles in the dynamic spectrum.
For this dataset, however, there is a compromise that needs to be made as larger windows result in biases in our scintillation timescale measurements, as $\Delta\tau_d$ changes rapidly across the orbital phase (as seen in \ref{fig:dynspec}).
We found that $10$\,min windows provided a sufficiently large number of scintiles and a sufficiently small fraction of the orbit to allow us to both reliably measure scintillation properties, but not be affected by intra-slice orbital variations.

To improve the reliability of error estimation for $\Delta\nu_d$ and $\Delta\tau_d$, we introduce white noise parameters, $F$ and $Q$, which account for epoch-to-epoch excess noise that could be systematic or astrophysical in origin\citep{Askew2023}. The uncertainty on the measurements is modified to be
\begin{eqnarray}
    \sigma_P = \sqrt{(\sigma_s F)^2 + \left( Q \left(\frac{\nu}{\nu_n}\right)^{\xi} \right) ^2},
    \label{eqn:sigma}
\end{eqnarray}
where $\sigma_P$ is the total uncertainty in the measurement, $\sigma_s$ is the measured uncertainty, $F$ is a scaling factor, and $Q$ is a frequency-dependent quadrature error, measured relative to a reference frequency $\nu_n$ (we used 1\,GHz) with a spectral exponent $\xi$.
These were modified from \citet{Askew2023} to account for the wider bandwidths of our observations ($\sim$2\,GHz, compared to 300\,MHz in \citet{Askew2023}). 

While trying to improve ACF fitting, we also explored other methodologies, which were not found to significantly affect the measurements or uncertainties.  
We considered an alternate method of weighting when fitting a model to the ACF, scaling the uncertainties on the ACF measurements to be proportional to the lag from the centre of the ACF.  
This was chosen to account for fewer measurements of the ACF at greater lags.
However, this only improved the precision of our measurements of the scintillation bandwidth by $\approx 2$\% (for UHF observations) and did not improve the systematic uncertainties described above.
To estimate the uncertainty of $\Delta\nu_d$ and $\Delta\tau_d$, we assume that the 2D-ACF samples are independent, which is a standard practice in scintillation analysis.
While the finite scintle effect accounts for correlations between lags from scintillation, other correlated noise sources could introduce correlations in the ACF. 
We have not considered the impact of this assumption.

\subsection{Spectral dependence of scintillation bandwidth}\label{chapt:alpha}

Scintillation is a strongly chromatic process. 
The degree of chromaticity depends on the structure of the inhomogeneous density fluctuations in the IISM \citep{Lambert1999}.
These inhomogeneities also impact other properties of the scintillation pattern, including the relationship between scintillation velocity and scintillation bandwidths and timescales \citep{Cordes1998, Lambert1999}.
Determining how scintillation bandwidth scales with frequency is crucial in establishing a scintillation-based orbital model that can be used to infer the distance to a pulsar.

We model the frequency dependence of the scintillation as a power law with exponent $\alpha^{\prime}$, such that
\begin{eqnarray}
    \Delta\nu_d \propto \nu^{\alpha^{\prime}}.
    \label{eqn:dnuVnu}
\end{eqnarray}

The value of the exponent depends on the underlying density fluctuations in the IISM. 
The electron density power spectrum, $P_{n_e}(q;z) \propto q^{\beta}$, is defined with a power law relationship with the exponent $\beta$ \citep{CPL1986, Rickett1990, Lambert1999}.
This exponent can be related to the spectral dependence of the scintillation as
\begin{eqnarray}
\alpha^{\prime} = 
  \begin{cases} 
      \frac{2 \beta}{\beta - 2} & \beta < 4, \\
      \frac{8}{6 - \beta} & \beta > 4, \\
   \end{cases}  
    \label{eqn:alpha}
\end{eqnarray}
\citep{Bhat2004}.
There is no known theoretical relationship between $\alpha^\prime$ and $\beta$ when $\alpha^{\prime} < 4$ \citep[e.g., see Figure 16 in][]{CPL1986}.

We measure the change in $\Delta\nu_d$ as a function of frequency with a global fit to $\alpha^{\prime}$ across all three observing bands.
To determine the scintillation bandwidth at each epoch, we fitted $\alpha^{\prime}$ using the relationship,
\begin{eqnarray}
\Delta\nu_d = \Delta\nu_{\textrm{d, 1GHz}} \left( \frac{\nu}{\nu_{\textrm{ref}}} \right)^{\alpha^{\prime}},
    \label{eqn:dnuGHz}
\end{eqnarray}
where $\nu_{\textrm{ref}} = 1$\,GHz.

\subsection{Scintillation Velocity}\label{chapt:ScintVelocity}

We can infer properties of the transverse motion along the LoS using our measurements from the dynamic spectra.
Central to this is the calculation of the scintillation velocity,
\begin{eqnarray}
    V_{\textrm{iss}} = \left[ \frac{A_{\textrm{iss}}}{\textrm{km}\,\textrm{s}^{-1}} \right] \left[ \frac{D}{\textrm{kpc}} \frac{\Delta\nu_d}{\textrm{MHz}}\right]^{1/2} \left[ \frac{\nu}{\textrm{GHz}} \frac{\Delta\tau_d}{\textrm{s}}\right]^{-1},
    \label{eq:viss}
\end{eqnarray}
where $A_{\textrm{iss}}$ is a constant that relates the screen geometry and density variations to the scintillation pattern speed \citep{Cordes1998}.
This geometric constant is defined to be
\begin{eqnarray}
    A_{\textrm{iss}} &&= \left( \frac{c}{4 \pi C_1} \right)^{1/2} \left( \frac{C_u}{C_1} \right)^{1/2}  \left(\frac{2 \left(1 - s\right)}{s}\right)^{1/2} \nonumber \\
     &&= A_{\textrm{iss}, 5/3, \mu} \,\, W_C \,\, W_{D, \textrm{ISS}},
	\label{eq:aiss}
\end{eqnarray}
where $s$ is the relative distance to a thin screen of plasma (where $s=1$ is the position of the Earth, and $s=0$ of the pulsar), $W_C$ and $W_{D,\textrm{ISS}}$ are variables that depend on the assumed spatial distribution and relative distance of the intervening plasma.
For every LoS one must consider what spatial and spectral models to assume to derive a value of $A_{\textrm{iss}}$.
When deriving a value of $A_{\textrm{iss}}$, implicit unit conversion is applied to have units of km\,s$^{-1}$ \citep{Cordes1998}.
The constant $C_1$ defines the Fourier relationship between the decorrelation bandwidth, $\Delta\nu_d$, and pulse-broadening timescale, $\tau_s$,
\begin{eqnarray}
    C_1 = 2 \pi \Delta\nu_d \tau_s
    \label{eq:C1}
\end{eqnarray}
For a uniform distribution of electrons in a Kolmogorov turbulent medium, $C_1 = 0.741$ \citep{Cordes1998, Lambert1999}.
However, the parameter $C_1$ depends on the geometric and spectral properties along the LoS.
These properties are described in greater detail in Section \ref{chapt:alphadiss}.

The methods for determining $C_1$ (and therefore $A_{\textrm{iss}}$) from scintillation have not been developed since \citet{Rickett2014}, which itself relies on the theory presented in \citet{Lambert1999} and \citet{Cordes1998}.
It is therefore not currently possible to calculate values $C_1$ when $\alpha^\prime < 4$ due to an unknown relationship between $\alpha^\prime$ and $\beta$. 
For completeness, we describe the methodology for determining $C_1$ and $D$ assuming $\alpha^\prime = 4.4$ in Appendix \ref{chapt:AppendixA}.

We can infer properties of the IISM and the orbit of the double pulsar by comparing $V_{\rm ISS}$ calculated from the scintillation time scale and bandwidth with a model: 
\begin{eqnarray}
   \textbf{V}_{\textrm{eff}}(s) &&= (1-s)(\textbf{V}_p + \textbf{V}_\mu) + s\textbf{V}_{\textrm{E}} - \textbf{V}_{\textrm{IISM}} \nonumber \\
    &&= \textbf{V}_{\rm kin} - \textbf{V}_{\textrm{IISM}},
	\label{eq:veff}
\end{eqnarray}
where $\textbf{V}_p$ is the transverse component of the orbital velocity of the pulsar, $\textbf{V}_\mu$ is the pulsar transverse space velocity, $\textbf{V}_{\textrm{E}}$ is the transverse velocity of the Earth, and $\textbf{V}_{\textrm{IISM}}$ is the IISM velocity \citep{Cordes1998, Askew2023}.
$\textbf{V}_{\textrm{IISM}}$ is parameterised by its relative velocity in right ascension (RA, $\alpha$) and declination (DEC, $\delta$): $V_{\textrm{IISM},\alpha,\delta}$.
We measure $\textbf{V}_p$, which importantly depends on $i$ and $\Omega$, to be between $\approx 25$ and $\approx 315$\,km\,s$^{-1}$ using the timing model from \citet{Kramer2021}.
$\textbf{V}_\mu$ has been held fixed using accurate measurements of the proper motion from \citep[$V_{P, \alpha}=-2.47(3)$ and $V_{P, \delta}=2.04(3)$][]{Kramer2021}.
$\textbf{V}_{\textrm{E}}$ and $\textbf{V}_p$ are calculated at the epoch of each measurement in the data and fixed at that value.

This velocity model assumes that a thin screen of plasma dominates the scattering along the LoS \citep{Sutton1971}.
When modelling $\textbf{V}_{\textrm{eff}}(s)$, it is possible to infer the fractional distance to the scattering screen, $s$, to a few per cent \citep{Reardon2020, Mall2022, Ocker2024, Reardon2024b}, which corresponds to a precision in physical precision of tens to hundreds of pc.

We consider two approaches fitting our model of $\textbf{V}_{\textrm{eff}}$ to the data.
For the first model (which we will refer to as the $\Delta\tau_d^{-1}$ model), we use the measured scintillation timescales and model the spatial scale,
\begin{eqnarray}
\frac{s_d}{\tau_d} = V_{\textrm{iss}}, 
\end{eqnarray}
where
\begin{eqnarray}
    s_d \equiv A_{\textrm{iss}} \sqrt{D \Delta\nu_d} \,\,\,\nu^{-1}
    \label{eq:spatialscale}
\end{eqnarray}
is the spatial scale, which is the distance over which intensity fluctuations caused by interference become uncorrelated \citep{Rickett1977, Armstrong1995, Cordes1998, Lambert1999, Rickett2014, Reardon2020}\footnote{This parameter is occasionally referred to as $l$ instead of $s_d$ and is not to be confused with the relative distance to the scattering screen $s$.}.
At each epoch, the spatial scale is measured relative to a reference frequency of $\nu_n= 1$\,GHz.
The spatial scale for different frequencies as the same epoch can be calculated as 
\begin{eqnarray}
    s_{d,n} = s_d \left( \frac{\nu}{\nu_n} \right)^{\gamma}.
    \label{eq:spatialscale_ellipse}
\end{eqnarray}

Using the second model (which we refer to as the $V_{\textrm{iss}}$ model), we derive values for $V_{\textrm{iss}}$ from our scintillation observations using Equation \ref{eq:viss}.
For this model, we scale the frequency dependence bandwidths $\alpha^{\prime}$ to account for the variations from epoch to epoch.
This model can be used to infer posterior distributions of $A_{\textrm{ISS}}$ and the distance to the double pulsar $D$, which are discussed in Section \ref{chapt:ResVissModel}.

We note that both models use the velocity of the LoS at the Solar System barycentre, which we define to be 
\begin{eqnarray}
    \textbf{V}_{\textrm{LoS}}(s) = \frac{\textbf{V}_{\textrm{eff}}(s)}{s},
	\label{eq:vlos}
\end{eqnarray}
This is important when comparing our results with other works, as we define the effective velocity at the Earth.
All of our models account for anisotropic scattering using methods described in Section \ref{chapt:AnisotropyMethod}.

\subsection{Velocity Modelling with Bayesian Inference}\label{chapt:VelcityModelling}

We use Bayesian modelling techniques described in \citet{Askew2023} to infer model parameters and their posterior probabilities.
We assume a Gaussian (logarithmic) likelihood.  
In the case of the $\Delta\tau_d^{-1}$ model, the likelihood has the form
\begin{eqnarray}
    \log\left(\mathcal{L} \right) = -\frac{N}{2}\log(2\pi\sigma_P^2)-\frac{1}{2\sigma_P^2} \sum_{n=1}^N \left[ \frac{s_{d,n}}{ \Delta\tau_{d, n}}- \textbf{V}_{\textrm{eff}, n}\right]^2,
    \label{eqn:norm_pdf}
\end{eqnarray}
where $\Delta\tau_{d,n}$ is the $n$th measurement of the scintillation timescale from $N$ total measurements, and $s_{d,n}$ is the corresponding inferred spatial scale for this $n$th measurement.
The modelled effective velocity for this measurement is $\textbf{V}_{\textrm{eff},n}$.   
The likelihood used for inference using measurement of $V_{\textrm{iss}}$ has a similar form.
In this work, unless explicitly stated, we use uniform priors across the entire (or reasonably broad) parameter spaces in our inferences.
When comparing models we can use Bayesian Evidence to select preferred models.

\subsection{Assessing Anisotropy}\label{chapt:AnisotropyMethod}

Scattering need not be isotropic.
Anisotropic scattering manifests as variations of the scatter-broadened image of the pulsar, approximated as an ellipse, rather than being azimuthally symmetric shape \citep{CPL1986, Coles2005, Rickett2014, Reardon2019}.
In the case of ellipsoidal anisotropy, the image of the pulsar is broadened non-axisymmetrically, and the anisotropy can be parametrized by an axial ratio $A_R$ and position angle $\zeta$ measured East of North of the major axis of the ellipse.
Other studies, including those outside pulsar astronomy, indicate anisotropic scattering being dominant along many lines of sight \citep{Desai2001, Chandran2002, Pen2012, Florinski2024}.

We also consider an extension to this simplest model for anisotropy.
This is motivated by a previous study of the double pulsar system \citep{Coles2005}, which used an ellipse to model the observed spatial variations in the scattering from pulsars A and B.
They found that while the model was a reasonable fit for the data, the ACF showed additional structure \citep[e.g., Figure 3 of][]{Coles2005}.

As an alternative to an ellipse, we model the scatter-broadened image as a sum of circular harmonics.
In a circular harmonic basis, the zeroth degree is a circle, the first is an ellipse, and the second (and higher) degrees are perturbations to that ellipse.
We consider perturbations up to the third order.
We can define harmonics up to this degree using six parameters, $\zeta$ and $A_R$ for the ellipse, $\Delta_2$ and $\zeta_2$ for the second-degree parameters, and $\Delta_3$ and $\zeta_3$ for the third degree.
In this notation, $\Delta_{2, 3}$ defines the amplitude of the harmonic and $\zeta_{2,3}$ the orientation of the perturbations.
The higher-order harmonics allow for a diversity of unique shapes that the IISM could produce due to diffractive or refractive effects in turbulent or intermittent plasma.
To remove the unlikely realisations of the circular harmonics with shapes that do not represent a physically motivated scattered-broadened image of the pulsar, we restricted the prior distributions for the amplitude perturbations $\Delta_{2, 3} \leq 1$.

This methodology fits nicely into the existing $\Delta\tau_d^{-1}$ model.
The perturbed ellipse can be used to calculate a modified spatial scale, $s_{\textrm{ell}} = s_d r_{\textrm{ell}}$, which depends on the radial position of the perturbed ellipse, $r_{\textrm{ell}}$).

\section{Results}\label{chapt:Results}

A summary of our results for the range of models considered is presented in Table \ref{tab:table_model}.
The data and model using the best-fitting parameters, for the $\Delta\tau_d^{-1}$ model are shown in Figure \ref{fig:taumodel}.
We find that a single screen at a relative distance of $s = 0.53^{+0.01}_{-0.02}$ dominates the scattering. 
For this screen, we measure moderate anisotropy with an axial ratio of $A_R = 2.4 \pm 0.2$ and an anisotropy angle $\zeta = 180^{\circ} \pm 2^{\circ}$ East of North.
This was supported with $\log({\textrm{BF}}) \approx 72$ compared to an isotropic model.
All of the figures in this section, unless otherwise stated, show the results inferred from this model.
We also find that this model breaks the degeneracy in the sense of the inclination angle, which pulsar timing is not always sensitive to, with our observations supporting an inclination angle $i> 90^\circ$. 

\begin{table}
	\centering
	\caption{ $\textbf{V}_{\textrm{eff}}$  inference. The results presented in this figure use the methods discussed in \ref{chapt:VelcityModelling} for the ($\Delta\tau_{d}^{-1}$) and ($V_{\textrm{ISS}}$) models. Both models support an inclination angle sense of $i > 90^{\circ}$, assuming the fixed magnitude of inclination angle of $i = 89.65^{\circ}$ from  \citet{Kramer2021}. The differences in $\xi$, $F$, and $Q$ are expected due to the different measurements for each model. $\gamma$ was not inferred within the $V_{\textrm{ISS}}$ model.}
	\label{tab:table_model}
	\begin{tabular}{l | c r}
		\hline
		\hline
        Parameter & $\Delta\tau_{d}^{-1}$ & $V_{\textrm{ISS}}$ \\
    	\hline
        $i$ ($^{\circ}$)& $>90$& $>90$\\
        \\
        $\Omega$ ($^{\circ}$) & $40_{-3}^{+3}$ & $52_{-3}^{+3}$\\
        \\
        $s$& $0.53_{-0.02}^{+0.01}$ & $0.60_{-0.01}^{+0.01}$\\
        \\
        $V_{\textrm{IISM},\alpha}$ (km\,s$^{-1}$) & $-22_{-1}^{+1}$ & $-26_{-3}^{+3}$\\
        \\
        $V_{\textrm{IISM},\delta}$ (km\,s$^{-1}$)& $36_{-3}^{+3}$ & $35_{-3}^{+3}$\\
        \\
        $A_R$ & $2.4_{-0.2}^{+0.2}$ & $2.0_{-0.2}^{+0.2}$\\
        \\
        $\zeta$ ($^{\circ}$) & $180_{-2}^{+2}$ & $169_{-3}^{+3}$\\
        \\
        $\gamma$ & $0.69_{-0.03}^{+0.03}$ & $-$ \\
        \\
        $\xi$ & $-1.7_{-0.6}^{+0.5}$ & $-6_{-2}^{+2}$\\
        \\
        $F$ & $2.30_{-0.08}^{+0.08}$ & $4.0_{-0.1}^{+0.1}$\\
        \\
        $Q$ (s$^{-1}$ / km\,s$^{-1}$) & $6.0^{+0.6}_{-0.7} \times 10^{-4}$ & $2.4_{-0.5}^{+0.7}$\\
		\hline
	\end{tabular}
\end{table}

\subsection{Spectral and Geometric line of sight properties}\label{chapt:alphadiss}

Understanding the IISM using scintillation observations can help answer two questions: (1) Where are the electrons along the LoS (i.e., where are the scattering regions)? and (2) How do the inhomogeneities in the IISM impact the scattering at different observing frequencies (i.e., what are the structures of the scattering regions)?
A common assumption is that the scattering is caused by electrons with density variations following a Kolmogorov power law ($\alpha^{\prime} = 4.4$, equivalent to $\beta = 11/3$) residing in geometrically thin screens \citep{Sutton1971, CPL1986, Rickett1990}.

A small change in $\alpha^{\prime}$ can result in a significantly different physical interpretation of a scattering medium.
Previous studies of the frequency dependence of diffractive scintillation have shown that spectral indices deviate significantly from those expected from Kolmogorov turbulence.
The majority of pulsars show power-law exponents ($\alpha^{\prime}$) that are shallower than what is expected from Kolmogorov, with the mean estimated to be $\alpha^{\prime} = 3.9 \pm 0.2$ in \citet{Bhat2004} and median to be $\alpha^{\prime}$ = $3.7^{+0.6}_{-0.7}$ in \citet{Krishnakumar2019}.
We have measured a shallow spectral index for the double pulsar where the scintillation decorrelation bandwidth scales with exponent $\alpha^{\prime}=3.61 \pm 0.05$.
The data strongly favour ($\log({\textrm{BF}})=119$) a shallower $\alpha^{\prime}$ than the value expected of Kolmogorov turbulence because of the large number of measurements involved in the inference (see Figure \ref{fig:alpha_prime}).
We note that this is the first measurement of the frequency dependence of scintillation of the double pulsar.

Several potential scattering geometries allow for spectral deviations from that expected from Kolmogorov turbulence.
For example, the electron density distribution may follow a power law spectrum with an `inner' or `outer' scale \citep{Lambert1999, Bhat2004}, or there may be a transverse truncation of the scattering screen \citep[i.e., scattering caused by thin sheets folding onto one another,][]{Pen2012, Jow2024}.

\begin{figure}
	\includegraphics[width=1\columnwidth]{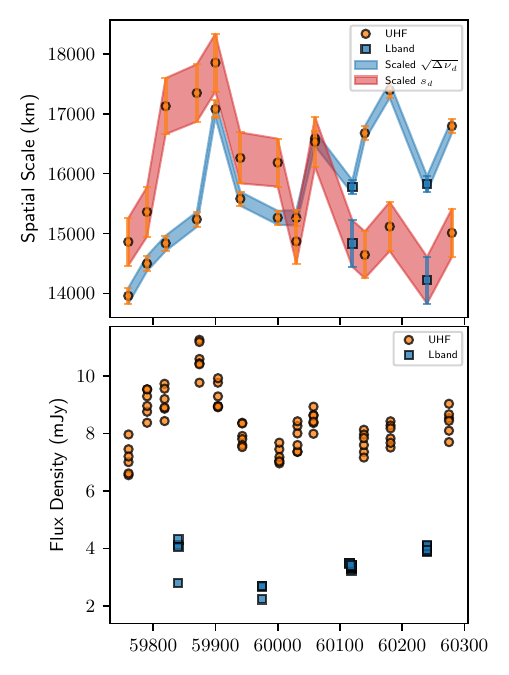}
    \caption{ Time series of spatial scale and flux density. The top panel shows the spatial scale variations using two models that depend on the epoch spatial scale (red) and the inferred spatial scale from the scintillation bandwidth measurements (blue), respectively. The middle panel shows the calibrated flux density taken for each observation. Measurements taken at UHF and L-band are shown with circles and squares, respectively. The flux density during late 2022 (MJD 59900) correlated ($\approx 0.66$) with the spatial scale from our scintillation measurements. The S-band data have not yet been flux-density calibrated.}
    \label{fig:Flux}
\end{figure}

We also note that the measurements of $\Delta\nu_{\textrm{d, 1GHz}}$ are correlated with flux density (see Figure \ref{fig:Flux}) with the correlation coefficient measured to be $\approx 0.65$.
This suggests refractive variations are impacting the diffractive scintillation measurements at each epoch \citep{Shannon2017}.

Both Kolmogorov and `square-law' ($\beta = \alpha^{\prime} = 4$) analytical models have been extensively examined in the literature \citep{Rickett1990, Lambert1999}.
We consider modifications to these models that include inner and outer scales defined to be $L_i = \kappa_i^{-1}$\,m, and $L_o = \kappa_o^{-1}$\,m, respectively \citep{Coles1987, Lambert1999}.
The inner scale model assumes a Kolmogorov medium that dissipates at a length scale $L_i$ at which kinetic energy begins heating the plasma \citep{Lambert1999} and suppressing the turbulent cascade. 
The outer scale model describes scattering from discrete objects (`blobs' of plasma) along the LoS with sharp boundaries, resulting in the square law distribution.
This can be used to describe scattering caused by \HII\, regions, plasma shock fronts at the boundaries of SNRs, and extreme scattering events \citep{Lambert1999, Lambert2000, Coles2015}.


\subsection{Anisotropy and IISM Variability}\label{chapt:anisresults}

\begin{figure*}
    \includegraphics[width=2.0\columnwidth]{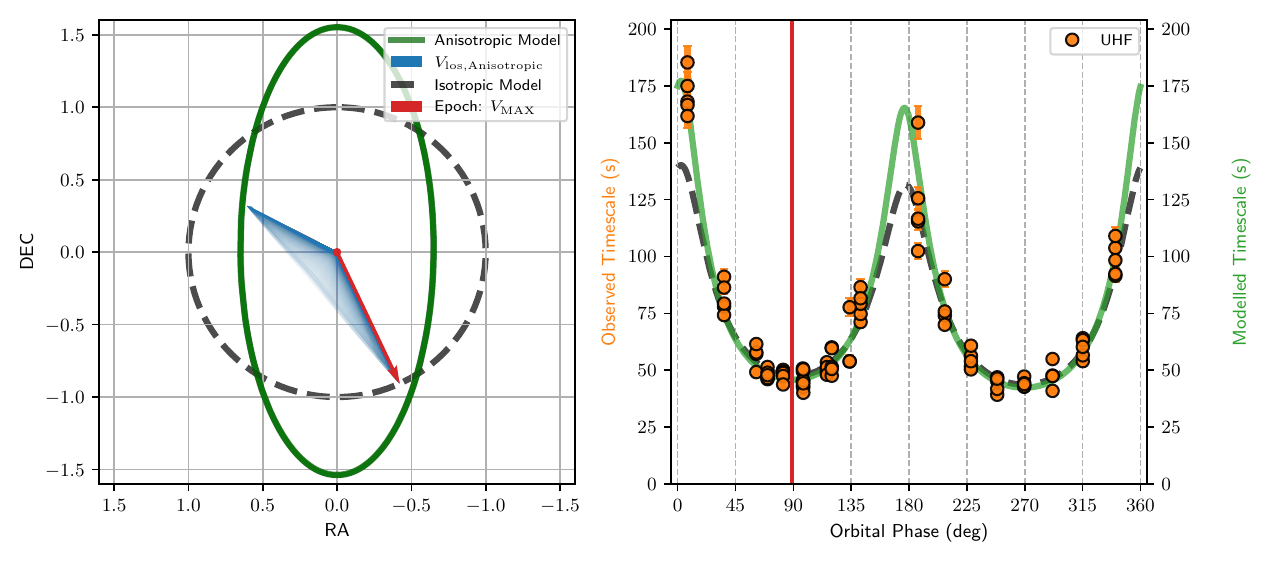} 
    \caption{ Comparison of models for anisotropic scattering. Left panel: the effective scattered image on the sky. The dashed black line indicates an isotropic scattering image on the sky. The dark green line shows the model using the best-fitting anisotropic model listed in Table \ref{tab:table_model}. The blue lines show the effective velocity (normalised relative to the maximum) at each sample from the model, and the red arrow corresponds to the velocity at the eclipse, where the transverse velocity is maximized. Right panel: Observed timescale measurements, plotted as circles, and the modelled timescale (solid line) shown against the orbital phase. The dashed black line is the same isotropic model from the left panel, and the red vertical line indicates the eclipse. The data and models are derived from the observations on UTC 2023-03-28.}
    \label{fig:anisotropy}
\end{figure*}

Using the model presented \ref{chapt:AnisotropyMethod}, we searched for distortions in the scatter-broadened image beyond a simple elliptical scattering disk.
These results favour an elliptical anisotropic model, which we use for the rest of our analysis.
The effect of this anisotropy on the spatial scale across the orbital phase is shown in Figure \ref{fig:anisotropy}, for a single UHF observation. 
The Figure also shows the isotropic model, which is a poor model for the observations.
We also searched for higher order perturbations using circular harmonics.
The amplitudes of the second and third circular harmonics $\Delta_{2}$ and $\Delta_{3}$ were found to be consistent with zero, with amplitudes constrained to be smaller than $3\%$ perturbations of the elliptical image.
Therefore, throughout the rest of the manuscript all models use the simpler anisotropic model.
However, future work may find that circular harmonics are necessary to describe stronger, more complex anisotropic scattering screens.

We also searched for epoch-to-epoch variations in the scattering screen properties.
This is motivated by the correlated, temporal variations in the spatial scale and flux density as shown in Figure \ref{fig:Flux}.
We focused our searches on the parameters that would be most likely affected by epoch-to-epoch variations: namely $A_R$, $\zeta$, and $s_d$.
We constrained the longitude of the ascending node, $\Omega$, to a Gaussian prior based on measurements and uncertainties derived from our $\Delta\tau_{d}^{-1}$ model.
We also assumed broad Gaussian priors for $V_{\textrm{IISM}, \alpha, \delta}$, centred on our measurements listed in Table \ref{tab:table_model}, with a standard deviation of 20\,km\,s$^{-1}$.
Other parameters were held constant as these were found not to correlate significantly with the aforementioned parameters.

\begin{figure*}
	\includegraphics[width=2\columnwidth]{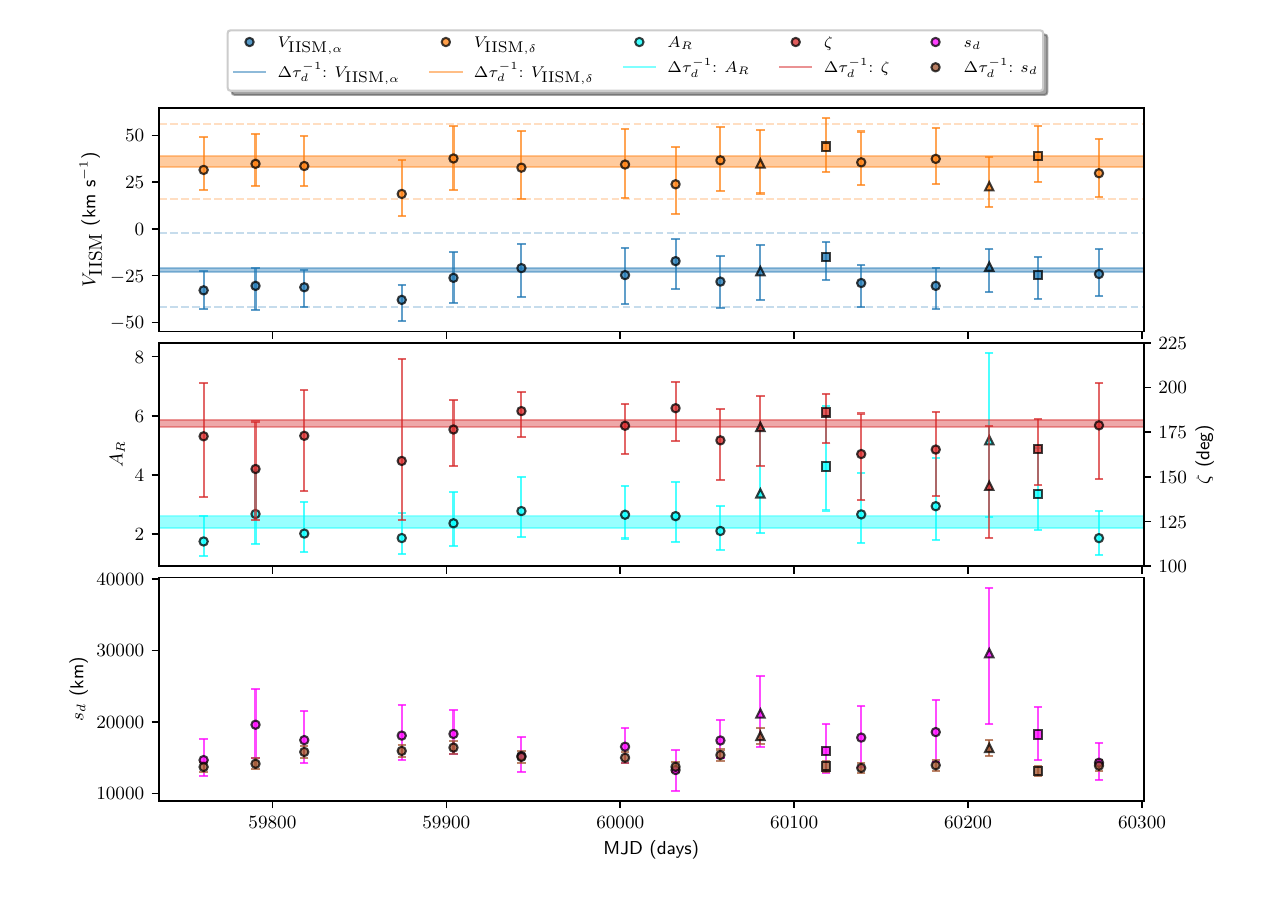}
    \caption{ Time series of scattering screen parameters. In each panel, the solid line shows the maximum likelihood parameter values from the $\Delta\tau_{d}^{-1}$ model. Each data point represents the maximum-likelihood parameters of the epoch-to-epoch model for UHF (circles), L-band (squares), and S-band (triangles). The dashed lines in the top panel show the prior range of the transverse IISM velocity parameters that were used to inform the epoch-to-epoch model. The middle panel shows the axial ratio and anisotropy angle, with scales indicated on the left and right y-axis, respectively. }
    \label{fig:EpochEpoch}
\end{figure*}

The results from this modelling are presented in Figure \ref{fig:EpochEpoch}.
We find that most of the parameters of the screen are stable between epochs and consistent with the $\Delta\tau_{d}^{-1}$ model.
The stability of the screen properties across a year is expected to be (and is) seen over longer timescales for other pulsars \citep{Reardon2020, McKee2022, Askew2023}.
We note that in the third panel of Figure \ref{fig:EpochEpoch}, the epochs at MJD $\sim$ 60080 (2023-05-16) and $\sim$ 60212 (2023-09-25) are S-band observations.
These epochs were found to have spatial scale measurements greater than the other observations.
As we have only two S-band observations, and neither is flux calibrated, it is difficult to determine more information about the sudden rise in spatial scale. 
Our results are in contrast to \citet{Rickett2014}, which reported significant variation in the anisotropy across different epochs.
This suggests that we could be observing scattering from a different screen from that modelled in \citet{Rickett2014}.  We consider this further in Section \ref{chapt:screendiss}.

\subsection{Distance Models}\label{chapt:ResVissModel}

We can use the properties of the scattering screen geometry to assess possible values of $A_{\textrm{ISS}}$ and what scattering screen geometries are consistent with published distance estimates to the double pulsar system.

\begin{figure*}
	\includegraphics[width=1.5\columnwidth]{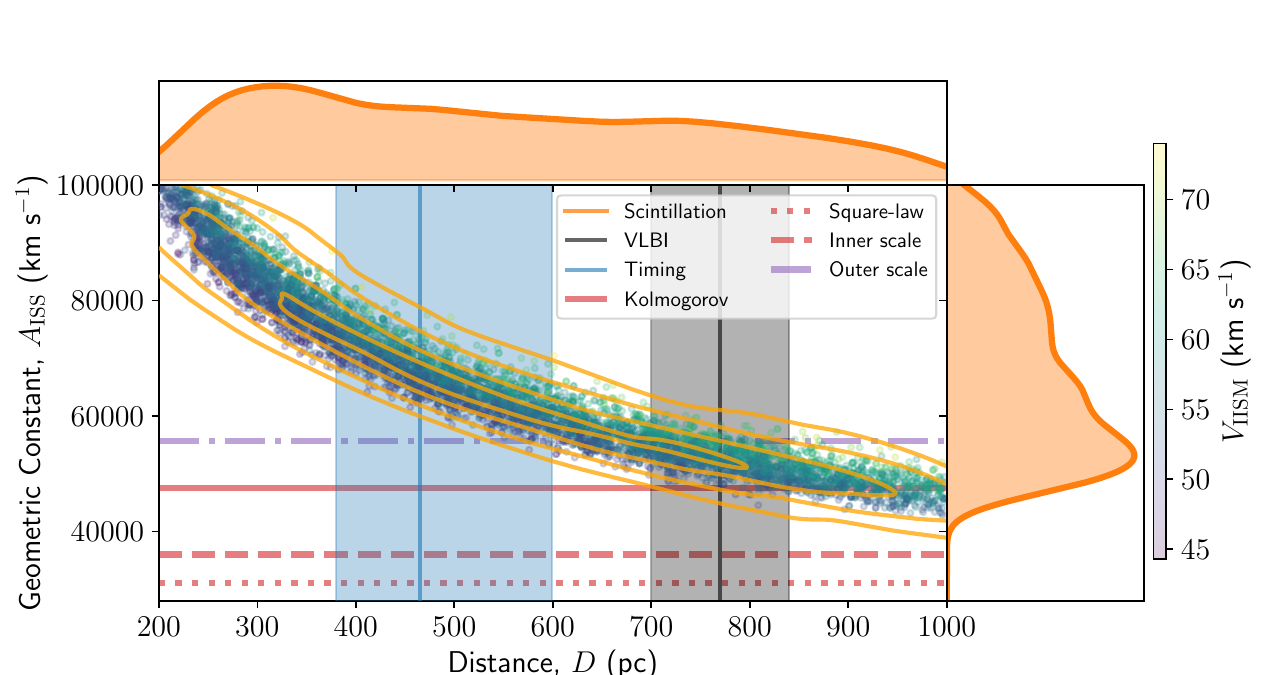}
    \caption{ Comparison of plausible distance and scattering geometries for the double pulsar. The orange contours show the $1-4\sigma$ confidence levels for the joint posterior distribution of $D$ and $A_{\textrm{ISS}}$. The vertical lines show the previous distance measurements from the timing (blue) and VLBI parallax (black), with the uncertainties shaded in the same colour to $1\sigma$ uncertainty. The horizontal lines show published calculations of $A_{\textrm{ISS}}$, assuming our inferred relative scattering screen distance. The colour bar represents the magnitude of the transverse scintillation component to the effective velocity.}
    \label{fig:distance}
\end{figure*}

As we cannot calculate $A_{\textrm{ISS}}$, we treat it as a parameter when assessing possible distances and scattering geometries. 
Figure \ref{fig:distance} shows the posterior distribution for $A_{\textrm{ISS}}$ and $D$ using the $V_{\textrm{ISS}}$ model.  
The distribution is broad because of the large degeneracy between $A_{\textrm{ISS}}$ and $D$.
To break this degeneracy, previous works have assumed a value for $A_{\textrm{ISS}}$ \citep{Ransom2004, Shapiro-Albert2020}.
In Figure \ref{fig:distance}, we show four possible values of $A_{\textrm{ISS}}$ published in \citet{Lambert1999}.
For each value, we assume a thin screen at a relative distance consistent with our measurements.
For the inner scale model, we assume Kolmogorov turbulence and an inner scale cut-off relative to the spatial scale at a wavenumber of $\kappa_i s_d = 10$.
For the outer scale model, we assume $\beta = 4$ and an outer scale cut-off relative to the spatial scale at a wavenumber of $\kappa_o s_d = 10^{-1}$.
By using Equation \ref{eq:viss} and assumed values for $s=0.5$,  $\Delta\tau_d=50\,$s, $\nu=1.3\,$GHz, $\Delta\nu_d=0.2\,$MHz, and assumed geometries $A_{\textrm{ISS}}$ (which also appear as horizontal lines in Figure \ref{fig:distance}), we can estimate the distance we might expect.
This is shown in the table below.

\begin{table}
	\centering
	\caption{ Assumed model parameters and geometries leading to distance estimates.}
	\label{tab:estimated_distance}
	\begin{tabular}{l | c c}
		\hline
		\hline
	    Turbulence Model & $A_{\textrm{ISS}}$ (km\,s$^{-1}$) & Distance (kpc) \\
    	\hline
        Kolmogorov  & $4.7\times10^4$ & 1.08 \\
        Inner-scale model & $3.6\times10^4$ & 1.83 \\
        Outer-scale model & $5.6\times10^4$ &  0.76 \\
        Square-law model & $3.1\times10^4$ & 2.47 \\
		\hline
	\end{tabular}
\end{table}

Assuming one of the distance measurements is correct (i.e., the distance to the double pulsar is $360$\,pc $\lesssim$ $D$ $\lesssim 840$\,pc), we constrain $A_{\textrm{ISS}}$ to be greater than $4.5 \times 10^{4}$\,km\,s$^{-1}$ and less than $8 \times 10^{4}$\,km\,s$^{-1}$.
Using Table \ref{tab:estimated_distance} and Figure \ref{fig:distance}, we can conclude that the timing distance is only consistent with the outer-scale model, and the VLBI distance is consistent with both the outer-scale and Kolmogorov models.

\section{Discussion}\label{chapt:Discussion}


\subsection{Previous scintillation studies of the double pulsar}\label{chapt:previousworksdiss}

The double pulsar has been studied using scintillation several times since its discovery \citep{Ransom2004, Coles2005, Stinebring2005, Rickett2014, Stock2024}, using various approaches and assumptions.
\citet{Ransom2004} assumed an isotropic scattering model to infer $V_{\textrm{ISS}}$.
A study of the scattering of radio waves from pulsar A travelling through the magnetosphere of pulsar B by \citet{Coles2005} found that anisotropic scattering ($A_R > 4$) was a better fit for their measured scintillation time scales than an isotropic model.
More recently, \citet{Rickett2014} studied the scintillation assuming an anisotropic model. 
Informed by scintillation timescale measurements, \citet{Rickett2014} used a similar method to our $\Delta\tau_d$ model.
These measurements were taken $\approx 17$\,yr ago, using the Green Bank Telescope.
They concluded that the anisotropy varied in both strength and position angle.  Interestingly, they inferred a screen at a distance of $s = 0.73 \pm 0.01$.
Since then \citet{Stock2024} has re-analysed the results and come to similar conclusions.

Some of our results disagree with those found in \citet{Rickett2014} and \citet{Stock2024}.
To understand the differences, we have partially re-analysed the data presented in \citet{Rickett2014} using our methodology and code.
We measured the fractional distance of the scattering screen to be $s = 0.70 \pm 0.01$ and the longitude of the ascending node to be $\Omega = 64^{\circ} \pm 8^{\circ}$, which is consistent with what they reported.
While the uncertainty on $s$ is consistent with their value, the uncertainty on $\Omega$ is a factor of four larger, which we attribute to our inclusion of terms to account for excess white noise.
Unlike \citet{Rickett2014}, we also searched for the longitude of periastron, $\omega$, to check for consistency with the known value from pulsar timing.
We compare the results from scintillation to the $\omega$ measured in \citet{Hu2022}, updated (using the measured rate of periastron advance) to an epoch close to the centre of our MeerKAT data set, $T_0 = 59976.7$\,MJD.
At this epoch, the longitude of periastron derived from pulsar timing is $\omega= 42.6169^{\circ} \pm 0.0006^{\circ}$ and at the reference epoch of the \citet{Rickett2014} data set ($T_0 = 53155.9$\,MJD)  the longitude derived from pulsar timing is  $\omega = 87.0309^{\circ} \pm 0.0005^{\circ}$ \citep{Hu2022}.
When fitting for $\omega$ in our analysis of the MeerKAT data set, we find a consistent value of $\omega = 42.5^{\circ} \pm 0.2^{\circ}$.
In contrast, when modelling the \citet{Rickett2014} data set, we find $\omega = 310^{\circ} \pm 30^{\circ}$, which is inconsistent with pulsar timing at the $3-\sigma$ level.
This suggests our data set and model may be more reliable than those presented in \citet{Rickett2014}.

Some of our results disagree with \citet{Rickett2014} and \citet{Stock2024} because of the choice of coordinate systems.   
We note that the definition of $\Omega$ in our analysis is consistent with the definition of \citet{Damour1992} and implemented in the scintillation code by \citet{Reardon2019, Scintools}.
However, the direction of the y-axis \citep[in the plane of the binary orbit, see Figure 2 of ][]{Stock2024} is inverted in \citet{Rickett2014} and \citet{Stock2024}  when compared to the \citet{Damour1992} definition.
This, in addition to the difference we see when measuring $\omega$ using \citet{Rickett2014} data, explains the differences we observe when comparing the orbital parameters of $i$ and $\Omega$ between \citet{Rickett2014}, the timing and our analysis.


We compare our inference on the orbital inclination to that derived using two other independent methods.
Firstly, it is also possible to infer the sense of the inclination angle by modelling the eclipses of pulsar A by pulsar B.
Using this method \citet{Lower2024} also found $i > 90^{\circ}$.
Secondly, the analysis of the polarization of the pulsar A pulse profile can be used to infer the system geometry using the rotating vector model \citep{RVM, Blaskiewicz1991} to measure the structure of the beam of pulsar A.
Using this model, \citet{Kramer2021b} estimate to be $i = 91.6^{\circ} \pm 0.1^{\circ}$.


In our modelling, we use the magnitude of the inclination angle from the timing model of \citet{Kramer2021} as a prior.
Our results support $i > 90^{\circ}$ (see Table \ref{tab:table_model}) with a log BF $\approx 21$.
This is therefore consistent with recent studies that find $i > 90^{\circ}$ \citet{Kramer2021b, Hu2022} and \citet{Lower2024} and inconsistent with \citet{Ransom2004} and \citet{Rickett2014}.

Our model of the relative screen distance (see Table \ref{tab:table_model}) is in strong tension with the results of \citet{Rickett2014}.
Given the difference between these observations ($\sim 17$\,yr), if a single screen was responsible for all of the observed scintillation, it would need to be at least $\mathcal{O}(100)$\,AU in projected transverse extent.
Multiple scattering screens along the LoS are not uncommon \citep{Mall2022, Ocker2024, Reardon2024b}.
These can often appear as scintillation arcs \citep{Stinebring2001} with different curvature values from distinct parabolic arcs in the secondary spectrum (the two dimensional spectrum of the dynamic spectrum). 
It is also not unlikely that the scattering screens can vary over $\sim$decades of observations \citep{Walker2022}.
Therefore, we conclude that we are probing a different area of the IISM along the LoS.

It is also possible to use the secondary spectrum to characterise the ISM and orbital dynamics of pulsars.
Indeed, previous work has also investigated the secondary spectrum in scintillation towards the double pulsar in \citet{Stinebring2005}.
The utility of using parabolic arcs depends on the sharpness of the arcs. 
At lower frequencies (UHF) the scintillation arcs are diffuse due to the relatively high strength of scattering and only modest anisotropy.
Therefore, \citet{Stinebring2005} conclude that observations at higher frequencies and with a greater signal-to-noise ratio would yield more promising results.
Our results agree. 
Work analysing the secondary spectrum of  GBT and MeerKAT data is ongoing (D.~Montalvo et al., in prep.).

\subsection{Distance to the double pulsar system}\label{chapt:distancediss}


Systematic errors may be impacting published distance measurements to pulsars \citep{Verbiest2012, Ding2023}.  
Some of the errors that impact pulsar timing include DM variations and intrinsic spin noise, which can be modelled \citep{Keith2013}, and mitigated \citep[][]{Goncharov2021}.
The high ecliptic latitude ($b=-51.2^{\circ}$) of the double pulsar system reduces the impact of the solar wind on DM variations but increases systematic errors and reduces the precision of a distance measurement \citep{Kramer2021}.

The measurement of a trigonometric parallax through VLBI can potentially be affected by refractive image wander caused by the turbulent IISM and the ionosphere of the Earth.
These effects in the current analysis of the double pulsar are negligible, however, this may change in the future, and other pulsars could benefit from a detailed understanding of the scattering \citep{Kramer2021}.
Refractive image wander can be caused by phase gradients in the IISM, which have been seen to change across the orbital and annual phase of the double pulsar \citep{Rickett2014}.
The strength of refractive image wander can be characterised by a refraction angle $\theta_r$.
The refractive scintillation for the double pulsar has a characteristic timescale on the order of months (discussed below).
As the Very Long Baseline Array (the instrument used for VLBI observations of the double pulsar) is in the northern hemisphere (and the double pulsar is in the southern sky), the array also has to point close to the horizon, potentially adding to the systematic uncertainty.
This is because the antennas in the array have to observe through large columns of the atmosphere and ionosphere of the Earth, impacting array calibration.

Our scintillation analysis can be used to critically examine the pulsar timing and VLBI-inferred distance measurements.
We can compare our estimate of refractive image wander to that previously inferred from DM variations.
We estimate the image wander from variations in the spatial scale.
We can attribute the variations in spatial scale to refractive effects because of the significant correlation between flux density variations and spatial scale.
We first estimate the Fresnel scale, $r_{\textrm{f}} = \sqrt{D_s/k}$, where $k = 2\pi / \lambda$ is the wavenumber.
We then calculate the strength of scattering, $u = r_{\textrm{f}} / s_d$, using the epoch-by-epoch spatial scale measurements.
Assuming Kolmogorov turbulence, the diffractive scattering angle and refractive image wander are $\theta_d = 2 \sqrt{2 \log{2}} \,\, \lambda / \left( \pi s_d \right)$ and $\theta_r = \theta_d u^{-1/3}$, respectively \citep{Rickett1990}.
In our observations, we find the RMS for $\theta_r$ to be $\approx$ 0.825\,mas at 1\,GHz.

This is notably larger than what was estimated by \citet{Kramer2021}.
From DM variations, they estimated the RMS variations in $\theta_r$ to be 0.177\,mas, at a reference frequency of  1.56\,GHz, However, the peak-to-peak variations in image wander were found to be as large as 0.8\,mas.
Our RMS  $\theta_r \approx 0.339$\,mas is much larger when scaled to this frequency, assuming $\theta_r \propto \nu^{-2}$. 
The differences are not surprising.
Approximately  $3$\,yr passed between the last of their measurements and the first of ours.
Alternatively, it could be the case that the ISM anisotropy is more pronounced in directions perpendicular to the pulsar proper motion, as we infer.  
In this case, the effects of image wander would have been underestimated by \citet{Kramer2021}.
If refractive image wander had annual or semi-annual variations, it could impact VLBI and pulsar timing measurements. 


Earlier in Section \ref{chapt:ResVissModel}, we introduced how a third independent distance measurement could potentially resolve tension between the timing and VLBI distances.
Here we determined that only the Kolmogorov and outer-scale models are consistent with this distance.
The frequency dependence of the scintillation bandwidth suggests the turbulence is not purely Kolmogorov.
A Kolmogorov model with an inner-scale or outer-scale component (different from those plotted in Figure \ref{fig:distance}) may describe the structure of the medium.
A similar argument can be applied to a square law medium with a different outer scale component. 
Further support for the VLBI distance comes from the association of the scattering screen with the Gum Nebula, which we discuss in the next section.

\subsection{Screen associations}\label{chapt:screendiss}

We next explore the possibility and implications of the observed scattering screen being associated with the Gum Nebula.
The Gum Nebula has several different speculated formation channels, including a supernova explosion, an \HII\, region, a wind-blown bubble, or possibly a combination thereof \citep{Reynolds1976A, Reynolds1976B, Woermann2001, Sushch2011, Purcell2015}.
The most common explanation,is that it is a SNR resultant from a supernova explosion occurring approximately  1\,Myr ago \citep{Reynolds1976A, Reynolds1976B}.
To investigate the progenitor of the nebula, \citet{Woermann2001} modelled the peculiar velocity of the runaway star $\zeta$ Puppis, which is thought to have gained its high velocity due to a kick after the supernova explosion of a binary companion.
They found that its path crosses the modelled centre of the asymmetric expansion of the neutral gas \citep{Reynolds1976B, Woermann2001}.
Assuming this scenario is correct, \citet{Woermann2001} date the age of supernova to be 1.5\,Myr ago.
For the subsequent analysis and text, we also assume the progenitor of the Gum Nebula is a SNR.

Estimations of the age of the supernova depend on the properties of the explosion and the ambient interstellar medium.
Using the dynamics of SNRs, \citet{Cioffi1988} estimated the ambient density of the IISM, $n_o$, that a SNR ``ploughed through'' to form what we observe today.
Estimations from \citet{Reynolds1976B} estimated this to be as low as $n_o\,\sim 0.1-0.2$\,cm$^{-3}$.
More recently \citet{Purcell2015} estimated it to be $0.2 < n_o < 0.9$\,cm$^{-3}$.
\citet{Sushch2011} estimated $n_o = 0.07$\,cm$^{-3}$, assuming an age of supernova 1.5\,Myr, kinetic energy of supernova explosion $E_{51} = $\,10$^{51}$\,erg, a distance to the Gum Nebula of 400\,pc and relative solar metallicity $\zeta_m = 1$.
The expansion velocity of the ionized gas, $V_s$, will also affect this inference, which was originally estimated to be 10-30\,km\,s$^{-1}$ \citep{Reynolds1976A}.

\begin{figure}
	\includegraphics[width=1\columnwidth]{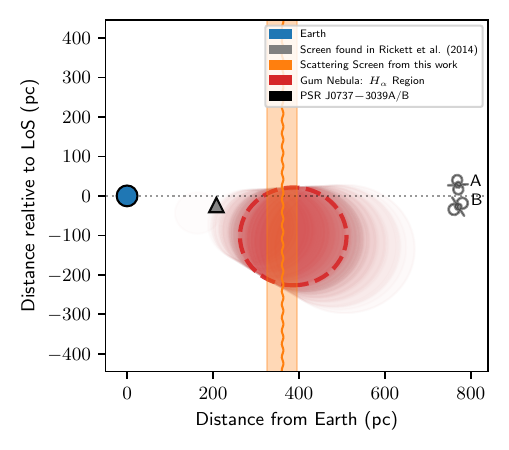} 
    \caption{ Model of the line of sight to the double pulsar. From left to right we show the Earth, the screen position measured by \citet{Rickett2014} (grey marker), our screen position (orange line), the Gum Nebula at a distance of $400 \pm 60$\,pc, with an inferred radius of 124\,pc at this distance (dashed red line) \citep{Sushch2011, Purcell2015, ONeill2024}, and the double pulsar J0737$-$3039A/B assuming the VLBI distance of 770\,pc. The transparent red circles are random draws are taken from the possible distance to the nebula, to show distance uncertainty.}
    \label{fig:ScreensCartoon}
\end{figure}

Recent work has modelled the Gum Nebula and local ionized gas regions, including the Vela supernova remnant and local bubble \citep{Sushch2011, ONeill2024}.
To simplify the model of the H$\alpha$ region, we assume a sphere at a distance of $400 \pm 60$\,pc with a radius of 124\,pc \citep{Sushch2011, ONeill2024}.
Figure \ref{fig:ScreensCartoon} shows a schematic representation of the LoS to the double pulsar, including the locations of the pulsar, the scattering screen, and the Gum Nebula.
The screen grazes the edge of the nebula.
The graphic also shows the position of the scattering screen measured by \citet{Rickett2014}.

An H$\alpha$ image of the ionized gas associated with the direction of the Gum Nebula is shown in Figure \ref{fig:Gum} from observations from the Southern H$\alpha$ Sky Survey Atlas \citep{Gaustad2001}.
The H$\alpha$ region is also represented in Figure \ref{fig:ScreensCartoon}, showing the overlap with the LoS to the double pulsar.
This indicates that, when assuming the VLBI distance to the double pulsar, our scattering screen is likely associated with the Gum Nebula.
The screen is therefore expected to be at a distance consistent with the outer edge of the nebula ($D_s = 360^{+30}_{-40}$\,pc).
It also suggests the distribution of material with the scattering screen is consistent having density fluctuations following an outer-scale model. 

We summarise evidence supporting plausible scattering screen models and distances in Table \ref{tab:ism_dist_comparison}.
If we assume the distance inferred from pulsar timing, the screen would be located at a distance $D_s = 220 \pm 30$\,pc from Earth so it would not be associated with the Gum Nebula.
Thus, the association of the scattering screen with the Gum Nebula provides additional support for the VLBI distance and an outer-scale model for the turbulence.
Throughout the remainder of the text, we assume the VLBI distance to the double pulsar for our analysis.

\begin{table}
	\centering
	\caption{ Comparison of ISM models and pulsar distance measurements.  An outer-scale scattering screen originating in the Gum Nebula and a VLBI distance to the double pulsar are the most consistent with our observations.}
	\label{tab:ism_dist_comparison}
	\begin{tabular}{l | c c}
		\hline
		\hline
	    Turbulence Model & Timing distance & VLBI distance \\
    	\hline
        Kolmogorov  & $\times$ & $\checkmark$ \\
        Outer-scale model & $\checkmark$ &  $\checkmark$ \\
        Inner-scale model & $\times$ & $\times$ \\
        Square-law model & $\times$ & $\times$ \\
        Gum Nebula & $\times$ & $\checkmark$ \\
		\hline
	\end{tabular}
\end{table}

\subsection{Applications of SNR association}\label{chapt:SNRmethods}

Using scattering screen velocities, it is possible to measure the expansion velocity, $V_s$, of the SNR.
This method assumes that the scattering screen is at the edge of a SNR where $V_s$ is the projected velocity given the measured transverse velocity $V_{\textrm{IISM}}$.
To calculate this, we used a vector from the centre of the Gum Nebula and used our measurement of the screen velocity $V_{\textrm{IISM}}$ to determine $V_s$.

The expansion of a SNR occurs in different stages \citep{Woltjer1972, Cioffi1988}, beginning with the Sedov-Taylor (ST) stage \citep{Taylor1950, Sedov1959}.
When the gas cools, a supernova remnant shell is formed at a time
\begin{eqnarray}
    t_{\textrm{sf}} = 3.61\times10^{4}\frac{E_{51}^{3/14}}{\zeta_{m}^{5/14} n_{o}^{4/7}},
    \label{eq:time_sf}
\end{eqnarray}
where $E_{51}$ is the kinetic energy of the supernova explosion in units of 10$^{51}$ erg, $\zeta_{m}$ is the metallicity (measured relative to solar metallicity)  that the shell ``snowploughs'' through the ambient IISM, of which the ambient density is  $n_o$ and expressed in units of cm$^{-3}$ \citep{Cioffi1988}.
The time at the end of the ST stage and the beginning of the pressure-driven snowplough (PDS) stage is given by $t_{\textrm{PDS}} = t_{\textrm{sf}}/e$, where $e$ is Euler's number.
We can then determine the age of a supernova $t$ using the normalised relative time $t_* = t/t_{\textrm{PDS}}$ \citep{Cioffi1988}:
\begin{eqnarray}
    t = 1.33 \times 10^{4} \frac{\, t_{*} E_{51}^{3/14}}{\zeta_{m}^{5/14} n_{o}^{4/7}} \,{\rm yr}.
    \label{eq:time_SN}
\end{eqnarray}
In the late stage of expansion of a SNR, the velocity and radius of expansion can be modelled as
\begin{eqnarray}
    v_s &&= 413\,n_{o}^{1/7} \zeta_{m}^{3/14} E_{51}^{1/14} \left( \frac{4}{3} t_{*} - \frac{1}{3} \right)^{-7/10} \nonumber \\ 
    &&= v_{\textrm{PDS}} \left( \frac{4}{3} t_{*} - \frac{1}{3} \right)^{-7/10} \,{\rm km\,s^{-1}},
    \label{eq:vel_SN}
\end{eqnarray}
and,
\begin{eqnarray}
    R_s &&= 14\, \frac{E_{51}^{2/7}}{\zeta_{m}^{1/7} n_{o}^{3/7}} \left( \frac{4}{3} t_{*} - \frac{1}{3} \right)^{3/10} \nonumber \\ 
    &&= R_{\textrm{PDS}} \left( \frac{4}{3} t_{*} - \frac{1}{3} \right)^{3/10} \,{\textrm{pc}},
    \label{eq:R_SN}
\end{eqnarray}
respectively \citep{Cioffi1988}.

Using these relationships we can express $t_*$ in terms of $R_s$, $V_s$, and $t$,
\begin{eqnarray}
    t_* = \left[ 4 - 1.18 \times 10^{6} \frac{R_s}{V_s t} \right]^{-1},
    \label{eq:t_star}
\end{eqnarray}
and then rearrange to solve for $E_{51}$ and $n_o$ in terms of $R_s$, $V_s$, $\zeta_m$, and $t$,
\begin{eqnarray}
    E_{51} = \frac{R_s}{14}^{2} \frac{V_s}{413}^{6} \zeta_m^{-1} \left( \frac{4}{3} t_* - \frac{1}{3} \right)^{18/5},
    \label{eq:E_51_SN}
\end{eqnarray}
\begin{eqnarray}
    n_o = \frac{R_s}{14}^{-1} \frac{V_s}{413}^{4} \zeta_m^{-1} \left( \frac{4}{3} t_* - \frac{1}{3} \right)^{31/10},
    \label{eq:no_SN}
\end{eqnarray}
which allows us to directly measure the energy of a supernova explosion and the ambient density of the IISM it passes through.

In Figure \ref{fig:Gum}, we show the projected transverse velocity of the screen on the sky, $V_{\textrm{scr}}= 29 \pm 4$\,km\,s$^{-1}$. 
Here, we have accounted for the differential rotation of the Galaxy at the screen location, based on code used in \citet{Shamohammadi2024}.
The direction of motion is only modestly askew from a radial direction from the purported centre of the nebula.
This could be caused by asymmetries in the ionized gas expanding into the ambient IISM if the screen is associated with the Gum Nebula.

In this case, the expansion rate of the ionized gas would be $V_{s} = 35 \pm 5$\,km\,s$^{-1}$.
As far as we are aware, this is the first time an expansion rate has been measured this way.
The expansion velocity of the ionised gas we measure is significantly larger than that expected for a \HII\, region which is thought to be \citep[$\sim 4$\,km\,s$^{-1}$,][]{Lasker1966}.
However, this value is consistent with the lower limit of the SNR expansion provided in \citet{Sridharan1992} and \citet{Woermann2000} of $V_{s} =$ 12\,km\,s$^{-1}$ and close to the initial estimates of 10-30\,km\,s$^{-1}$ \citep{Reynolds1976A}.

Our inferred velocity is also a factor of a few larger than the expected mean of the (turbulent) plasma velocity in the IISM $|V_{\textrm{IISM}}|=10$\,km\,s$^{-1}$ \citep{Goldreich1995}.
This suggests that there are shocks in the ionized shell of the Gum Nebula, resulting in gas travelling at supersonic speeds, or that the shell is associated with a blast wave such as from a supernova explosion.
While towards some pulsars, screen velocities \citep{Sprenger2022, Mall2022, Liu2023, Wu2024} are lower and consistent with thermal velocity, several works find screen velocities significantly greater than the mean plasma velocity \citep{Reardon2020, McKee2022, Walker2022, Askew2023, Reardon2024b}, like our measurements.

\begin{figure*}
	\includegraphics[width=2\columnwidth]{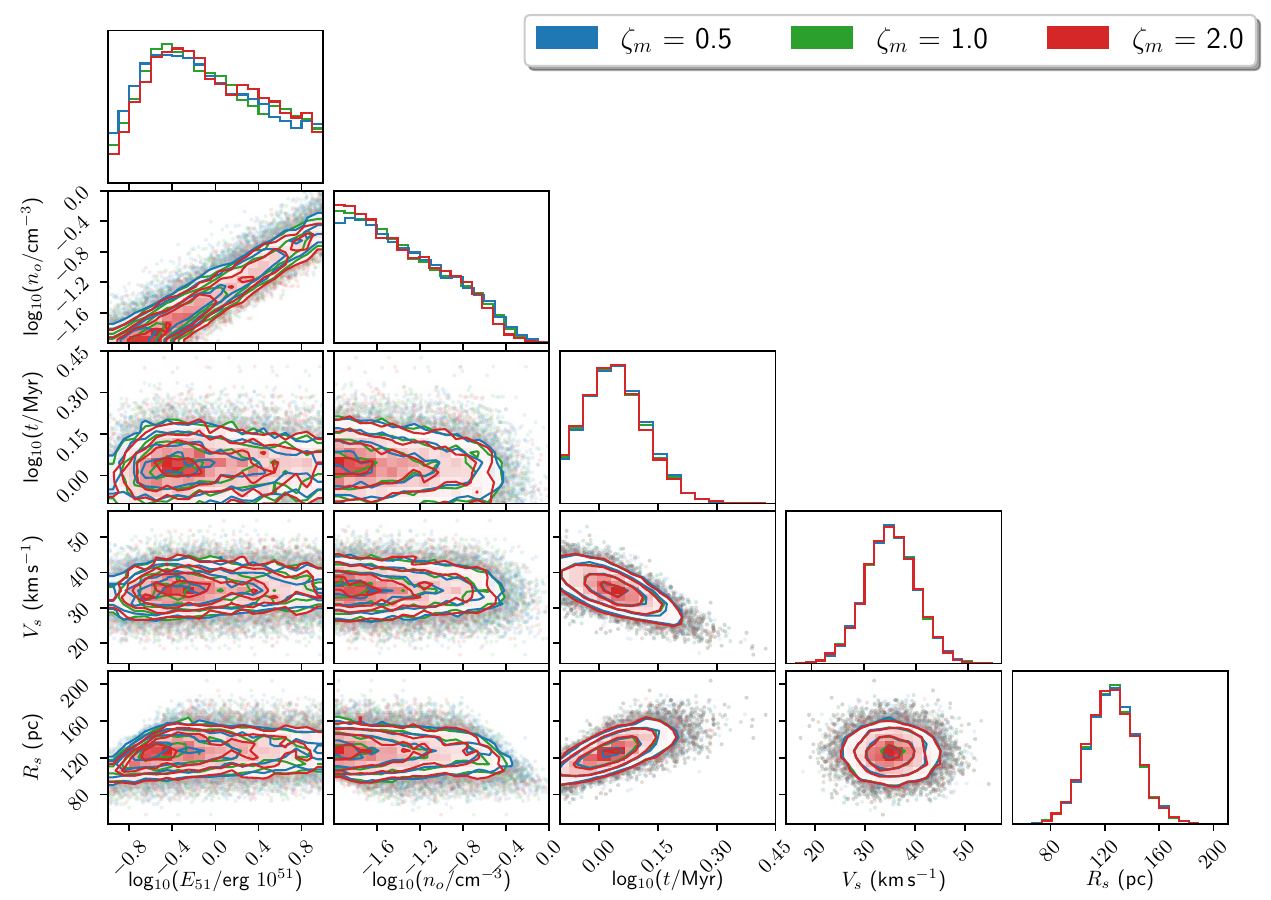} 
    \caption{ Two-dimensional posterior distributions for supernova scenario. Different metallicities (measured relative to solar metallicity) are shown in blue ($\zeta_m = 0.5$), green ($\zeta_m = 1.0$), and red ($\zeta_m = 2.0$). The normalised one-dimensional distributions appear to be weakly dependent on $\zeta_m$.}
    \label{fig:SNR_1}
\end{figure*}

We applied these results to the dynamics of SNR using equations \ref{eq:time_sf}-\ref{eq:no_SN}.
We used our estimates of $V_s$ and $R_s$ and their  1$\sigma$ uncertainties.
We limit the possible values for $E_{51}$, $n_o$, and $t$ to physical ranges of $10^{50}$ to $10^{52}$\,erg, $0.01$ to $1$\,cm$^{-3}$, and $0.5$ to $3.2$\,Myr, respectively.
The results can be visualised in the one- and two-dimensional marginal posterior distributions shown in Figure \ref{fig:SNR_1}.

Using these solutions to the Equations \ref{eq:time_sf}-\ref{eq:no_SN}, we find that the degree of metallicity has a negligible impact on the results, as expected.
Additionally, if we use the published estimates for $n_o$ \citep{Purcell2015}, we find these results are consistent with a supernova that is a few times more energetic than previously predicted.
We measure the age of the supernova to be $t = 1.1 \pm 0.1$\,Myr.
An expected age of the SNR of $t \approx 1$\,Myr is consistent with \citet{Reynolds1976B} given our expansion velocity and radius of the Gum Nebula.
Our estimate of the age of the supernova means the association with $\zeta$ Puppis with the formation of the Gum Nebula is less likely.
 
\section{Conclusions}\label{chapt:Conclusions}

The double pulsar has significantly advanced several aspects of fundamental physics.
These tests have been made possible because of a concentrated effort to understand and model the times of arrival from both pulsars A and B.
One of these tests is the measurement of the orbital decay due to gravitational wave emission.
Much like the Hulse-Taylor binary \citep{Hulse1975}, an accurate measurement of the distance is essential for this test and future precise tests of GR \citep{Kramer2021, Hu2022}.

We have developed a model of the scintillation, exploring its frequency dependence and measuring its spatial scale.
We find that the observed scintillation is consistent with originating in a single dominant anisotropic scattering screen located at the edge of the Gum nebula.
We have constrained the sense of the inclination angle of and measured the longitude of the ascending node of the orbit.

To accomplish this, we considered several possible models for the IISM and effects that could impact our analysis of diffractive scintillation patterns.
Our data supports a scattering screen with an outer scale.
Such screens are expected to be observed towards \HII\, regions or collections of plasma with steep edges, such as SNRs like the Gum Nebula.

While exploring possible associations along the LoS, we discovered that the scattering screen grazes the edge of the Gum Nebula H$\alpha$ emission.
We have measured the expansion of the ionized gas of the Gum Nebula to be $V_s = 35 \pm 5$\,km\,s$^{-1}$.
This was used to estimate the properties of the SNR, including the ambient density of the IISM the SNR ploughed through, the energy of the supernova explosion, and the relative solar metallicity of the ambient IISM.
In a novel way, using scintillation, we place limits on these physical properties of the IISM and find the age of the SNR to be $t \approx 1$\,Myr.

Any apparent tension between the pulsar timing and VLBI parallax distance can be solved by introducing a third independent measurement.
Using scintillation, we have provided independent support that VLBI measurements provide a more reliable distance to the double pulsar than those presently achieved through pulsar timing.
In the future it might be possible to use scintillation to make an independent distance measurement with improved theoretical modelling of the relationship between scintillation time scales, bandwidths, and velocities. 

Future work may determine $A_{\textrm{ISS}}$ when scintillation shows relatively weaker chromaticity, like we have observed.
This will allow for independent distance measurements to sources like the double pulsar.

If the distance is well constrained, scintillation observations can be used to test models of the IISM structure and to probe the properties of intervening objects along the line of sight.
Pulsars that are good targets for such a study are the binary millisecond pulsars PSRs J0437$-$4715, J1614$-$2230, J1643$-$1224, or J2222$-$0137 \citep{Guo2021, Mall2022, Shamohammadi2024, Reardon2024}.
This would help us develop an understanding of our Galaxy and the fundamental properties of scattering, which can impact precision pulsar timing.
However, if the distance is unknown and theory is developed to determine $A_{\textrm{ISS}}$ for a wider range of sightlines, a detailed study using scintillation could provide an independent measurement of distance for pulsars that show pronounced scintillation in the decimeter wavelength range, such as PSRs J0613$-$0200, J0614$-$3329, or J1757$-$5322 \citep{Gitika2023, Shamohammadi2024}.
Appendix \ref{chapt:AppendixA} presents a generalised method of determining the distances to pulsars using scintillation.

The double pulsar system will continue to provide improved tests of general relativity.
When it precesses back into the line of sight, scintillometry of the B-pulsar may be possible with the improved sensitivity of MeerKAT or the Square Kilometre Array. 

\section*{Acknowledgements}

We thank the referee for their valuable feedback, which has improved the structure and quality of the manuscript.
We acknowledge the Wurundjeri People, the traditional owners of the land on which Swinburne University of Technology (Hawthorn) is located.
This work was performed on the OzSTAR national facility at Swinburne University of Technology.
The OzSTAR program receives funding in part from the Astronomy National Collaborative Research Infrastructure Strategy (NCRIS) allocation provided by the Australian Government, and from the Victorian Higher Education State Investment Fund (VHESIF) provided by the Victorian Government.
The MeerKAT telescope is operated by the South African Radio Astronomy Observatory (SARAO), which is a facility of the National Research Foundation, an agency of the Department of Science and Innovation.
SARAO acknowledges the ongoing advice and calibration of GPS systems by the National Metrology Institute of South Africa (NMISA) and the time space reference systems department department of the Paris Observatory.
PTUSE was developed with support from the Australian SKA Office and Swinburne University of Technology. 
Parts of this work were funded through the ARC Centre of Excellence for Gravitational Wave Discovery (CE170100004 and CE230100016).
RMS acknowledges support through ARC Future Fellowship FT190100155.
This work is supported by the Max-Planck Society as part of the ‘LEGACY’ collaboration with
the Chinese Academy of Sciences on low-frequency gravitational wave astronomy.
This research has made use of NASA’s Astrophysics Data System and software packages, including: \texttt{MATPLOTLIB} \citep{Matplotlib}, \texttt{SCIPY} \citep{Scipy}, \texttt{ASTROPY} \citep{Astropy}, \texttt{NUMPY} \citep{Numpy}, \texttt{BILBY} \citep{Bilby}.

\section*{Data Availability}

Data and data products are available upon reasonable request to the corresponding author. 
 

\bibliographystyle{mnras}
\bibliography{example} 

\appendix
\section{Distance Measurements to Pulsars using Scintillation Velocities}\label{chapt:AppendixA}

Here, we describe the information and methods required to determine an independent distance to a pulsar using scintillation.
The modelling follows the Bayesian procedure outlined in Section \ref{chapt:VelcityModelling}. 
The required measurements include the scintillation bandwidth, $\Delta\nu_d$, timescale, $\Delta\tau_d$, phase gradient, $\phi$, and an estimation of $\beta$ from $\alpha^{\prime}$.
These can be extracted from the dynamic spectra of pulsar observations.

Following \citet{Lambert1999}, it is possible to calculate the uncertainty relation constant $C_1$.
This can be done by measuring the decorrelation bandwidth, $\Delta\nu_d$, and the scatter broadening timescale $\tau_s$.
One way to determine these properties is analytically by calculating the two-frequency plane-wave diffractive field coherence function, $\Gamma_D$, using, e.g., equation 59 in \citet{Lambert1999}.
This describes the scattering geometry and can be applied to spherical, plane waves, or those that propagate through an infinitely thin screen.
This can be extended to account for anisotropy in a thin screen \citep[][Equation A2]{Rickett2014}.
Within the \texttt{SCINTOOLS} package, we calculate the ACF of the electric field $\Gamma_D$ using
\begin{eqnarray}
    \Gamma_D \left( \mathbf{\sigma}, D_s; \nu, \Delta\nu \right) &&= \frac{\textrm{dsp}^2}{2 \pi i \Delta\nu_n} \int^{\infty}_{-\infty}\int^{\infty}_{-\infty} \exp{\left[-\frac{1}{2} D_\phi \left( \mathbf{\sigma}^{\prime}; \nu \right) \right]}  \nonumber \\
    &&\times \exp{\left[\frac{i}{2 \Delta\nu_n} \left| \sigma - \sigma^\prime \right|^2 \right]}d^2\sigma^\prime,
    \label{eq:Gamma_D}
\end{eqnarray}
where dsp is a parameter that describes the resolution of the ACF and
\begin{eqnarray}
    \Delta\nu_n = \frac{2 \pi \Delta\nu_d \nu^2 s_d^2}{D_s c}
    \label{eq:freqlag}
\end{eqnarray}
defines the value of each frequency lag within our calculations \citep{Scintools}.

This removes dependence on the distance, relative screen distance ($D_s$), scintillation bandwidth, and observing frequency.
We compute the ACF of the scattered electric field by solving equation \ref{eq:Gamma_D} at a grid of frequency and time lags.
We calculate the ACF of the intensity by multiplying $\Gamma_D$ by its complex conjugate.
We then measure $\Delta\nu_d$ and $\tau_s$ as the frequency lag at which the correlation falls to one-half and the time lag at which it falls to 1/$e$ of the peak value \citep{Lambert1999}.
These are then used to determine $C_1$ using Equation \ref{eq:C1}.
Code to calculate $C_1$ can be made available upon request.

Several parameters impact the form of $\Gamma_D$.
These include the spectral exponent, $\beta$, the axial ratio of anisotropy, $A_R$, the angle between the angle of anisotropy $\zeta$ and the direction of effective velocity $\textbf{V}_{\textrm{eff}}(s)$, $\psi$, the normalised phase gradient, $\phi_n$, and the angle between the phase gradient and the direction of $\textbf{V}_{\textrm{eff}}(s)$, $\theta_{\phi}$.
These parameters can be incorporated into codes that numerically integrate Equation \ref{eq:Gamma_D} and model scintillation velocities.

For pulsars, particularly those in binary systems, it may not be appropriate to determine a single value of $A_{\textrm{ISS}}$. 
For example, \citet{Rickett2014} showed that the phase gradient of the IISM varied over the binary period of the double pulsar.
It is often difficult to accurately measure $\psi$, $\phi_n$, and $\theta_{\phi}$ using scintillation measurements \citep{Rickett2014, Reardon2023}, so their broad uncertainties need to be accounted for in the analysis. 

It is therefore important to provide posterior distributions (or at least uncertainties) to constrain the parameters $\beta$, $A_R$, $\zeta$, $\psi$, $\phi_n$, and $\theta_{\phi}$ when quantitatively assessing the geometric structure of scattering and the numerical value of $C_1$ and its uncertainty.
$A_{\rm ISS}$ can be calculated from $C_1$ using Equation \ref{eq:aiss}.

This value of $A_{\textrm{iss}}$ can then be inputted as a prior into the model for $\textbf{V}_{\textrm{eff}}(s)$, which depends on $D$ and $s$, among other parameters.
Therefore, to determine $D$ using scintillation the measurements, it is necessary to model $\textbf{V}_{\textrm{eff}}(s)$.
This model can be fitted to an effective velocity inferred from the dynamic spectrum, by sampling $\Delta\nu_d$ and $\Delta\tau_d$ across annual and orbital phases (if the pulsar is in a binary).

Alternatively, we can conduct this analysis by using the relationship between the spatial scale and scintillation bandwidth,
\begin{eqnarray}
    s_d &&= A_{\textrm{iss}} \sqrt{D \Delta\nu_d} \nonumber \\ 
        &&= \kappa \sqrt{\Delta\nu_d},
    \label{eq:kappa}
\end{eqnarray}
where $\kappa$ is the generalized constant of proportionality between $s_d$ and $\Delta\nu_d$.
Then we can solve for $D$ using
\begin{eqnarray}
    D = \frac{\kappa^2}{A_{\textrm{iss}}^2}.
    \label{eq:distance}
\end{eqnarray}

\begin{figure}
	\includegraphics[width=1\columnwidth]{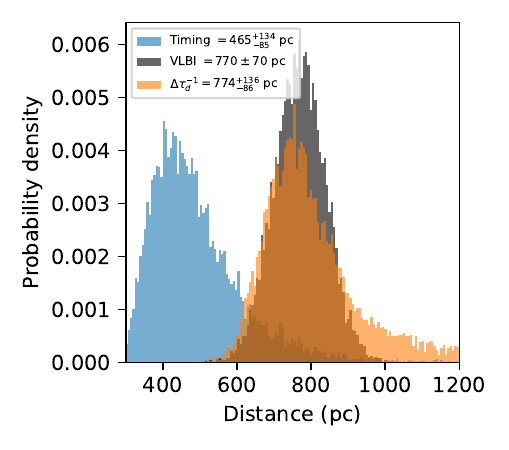}
    \caption{ Distance measurements for the double pulsar system. The VLBI (black) and timing (blue) distributions were calculated from \citet{Kramer2021}. The scintillation distribution (orange) was calculated assuming Kolmogorov turbulence.}
    \label{fig:distance_tau}
\end{figure}

We have measured the distance this way for the double pulsar when assuming $\alpha^{\prime} = 4.4$ (Kolmogorov turbulence) as seen in Figure \ref{fig:distance_tau}.
This distance agrees with the VLBI distance from \citet{Kramer2021}.
This demonstrates that scintillometry can provide comparable uncertainty to distance measurements from timing.

\section{Data Quality}\label{chapt:AppendixB}

\begin{figure}
	\includegraphics[width=1\columnwidth]{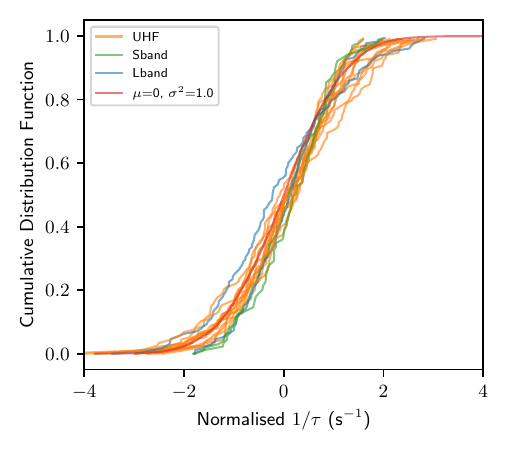}
    \caption{ Cumulative distribution of inverse scintillation timescale measurements. We compare these to the normalised residuals from Figure \ref{fig:taumodel}. Each frequency band is shown using a different colour. They are all consistent with a normal distribution with a variance equal to unity.}
    \label{fig:Tests}
\end{figure}

To test how well our model describes our data, we assessed the Gaussianity of the residuals.
The residuals are the difference between the data and model predictions, for scintillation time scales and velocities estimates.
We did this using the cumulative distribution function shown in Figure \ref{fig:Tests}.
We expect the normalised residuals (the residuals divided by the uncertainties modified by noise parameters $F$ and $Q$) to to have a Gaussian distribution with zero mean and unit standard deviation.
The expectation of such distribution is plotted in red, while the UHF, L-band and S-band measurements are plotted in orange, blue, and green, respectively.
We undertook this test for both the $\Delta\tau_{d}^{-1}$ and $V_{\textrm{ISS}}$ models.  
In the case of the $V_{\textrm{ISS}}$ model, the UHF data were consistent with the Gaussian distribution, but the L-band and S-band data were not, which is why we only used the UHF data in subsequent analysis. 
In addition to this, we computed the Anderson Darling statistic \citep[ADS;][]{AndersonDarling1954}, which can also be used to assess the Gaussianity of a data set.
We measured a statistic of ADS = 0.366 for the $\Delta\tau_d^{-1}$ model.
This was found to be smaller than the critical value for a normal distribution, suggesting that the distribution is consistent with a Gaussian distribution using this statistic as well.

\bsp	
\label{lastpage}
\end{document}